\documentclass[aip, reprint, twocolumn, floatfix, amsmath, amsfonts, amssymb]{revtex4-1}

\usepackage{bm}
\usepackage{graphicx}
\usepackage{xcolor}
\usepackage[colorlinks=true,linkcolor=blue,citecolor=blue,urlcolor=blue]{hyperref}

\newcommand{\Nhat}{\bm{\mathcal N}}
\newcommand{\Fock}{\hat F}
\newcommand{\Ham}{\hat H}
\newcommand{\wcis}{\omega_{\mathrm{CIS}}}
\newcommand{\wzero}{\omega^{(0)}}
\newcommand{\gh}{\bm{\gamma}_{\rm h}}
\newcommand{\gp}{\bm{\gamma}_{\rm p}}

\begin{document}

\title{State-specific perturbation theory for configuration interaction singles:
Theoretical analysis, benchmarks, and improvement with size-intensive level shifts}

\author{Takashi Tsuchimochi}
\email{tsuchimochi@gmail.com}
\affiliation{College of Engineering, Shibaura Institute of Technology,
3-7-5 Toyosu, Koto-ku, Tokyo 135-8548 Japan}

\begin{abstract}

We recently proposed a state-specific second-order perturbation theory based on the configuration interaction singles (CIS) state, termed CIS(2). Here, we clarify its theoretical connections, assess the accuracy of its excitation energies in greater detail, and develop size-intensive level shifts to improve both accuracy and convergence of its amplitude equations. We first derive the MP2 structure of its working equations and show that our previously introduced Hybrid partition recovers CIS(D) when the ground-state Fock operator is used and the residual Fock action on the CIS reference is neglected. For this partition, an expanded benchmark shows that using the CIS-state Fock operator improves Rydberg and especially charge-transfer accuracy relative to CIS(D), while slightly worsening valence accuracy. We analyze how the hole and particle densities contribute to these changes in excitation energy, which differ among the classes. We then introduce size-intensive level shifts that preserve the ground-state correlation contribution of noninteracting fragments while leaving the ground-state MP2 energy unchanged. With appropriately calibrated parameters, these shifts improve overall accuracy and substantially accelerate convergence without increasing the formal scaling of the amplitude solve. Different optimal parameters for valence and Rydberg states reveal the limitations of a common parameter. The analysis further shows that shifting the contracted-triples block contributes markedly to the excitation-energy correction in addition to accelerating convergence.

\end{abstract}

\maketitle

\section{Introduction}
\label{sec:intro}

An accurate yet affordable description of electronically excited states remains a central challenge in electronic structure theory.\cite{Dreuw2005Review,Gonzalez2012Review,Lischka2018Review} Configuration interaction singles (CIS)~\cite{ForesmanHeadGordon1992} provides the simplest wave-function description of excited states and is size-intensive;\cite{Dreuw2005Review,Bartlett2007CCReview} however, it lacks dynamical correlation and therefore systematically overestimates excitation energies, particularly for charge-transfer states.\cite{Subotnik2011CISCT} The most widely used remedy is CIS(D),~\cite{HeadGordon1994CISD,HeadGordon1995CISD} which adds a noniterative second-order correction to the CIS excitation energy; quasidegenerate and spin-component-scaled variants have also been developed.\cite{HeadGordonOumiMaurice1999CISDn,Grimme2007SCSCISD,RheeHeadGordon2007,Grimme2012SCSReview} CIS(D) was derived as a perturbative approximation to coupled-cluster response theory.\cite{HeadGordon1994CISD,StantonBartlett1993} Other second-order approaches, including CC2~\cite{Christiansen1995CC2,Hattig2000RICC2,Izsak2020CCReview} and ADC(2),~\cite{Schirmer1982ADC,Dreuw2015ADC} obtain excitation energies from response or propagator equations built on the ground state.

In our previous work,~\cite{Tsuchimochi2026paper1} we proposed CIS(2), a state-specific second-order perturbation theory in which the CIS state itself serves as the zeroth-order wave function and the zeroth-order Hamiltonian is constructed from the generalized Fock operator of that state. The first-order space consists of Hartree--Fock (HF)-parent doubles and contracted triples, the latter of which are generated by applying double excitations to the CIS state. The formulation does not require canonical HF orbitals and can therefore use orbitals optimized for CIS-type references, including excited-state mean-field theory (ESMF)~\cite{SheaNeuscamman2018ESMF,SheaGwinNeuscamman2020,Hardikar2020ESMFSCF,Zhao2020ESMF} and its state-averaged counterpart, SACIS.\cite{Tsuchimochi2026SACIS,Tsuchimochi2026SACISgrad} Its reference-based construction also provides a starting point for extensions to spin-projected CIS~\cite{Tsuchimochi2015SFCIS,Tsuchimochi2015TDPHF} and double CIS.\cite{Tsuchimochi2024DCIS}

In our previous study, the Canonical partition of CIS(2), which uses the CIS-state Fock reference gap, $\wzero$, systematically underestimated excitation energies, with a mean error (ME) of $-0.29$~eV for the 100 states of the QUEST\#1\cite{Loos2018QUEST1} benchmark.\cite{Tsuchimochi2026paper1} Replacing this gap by the CIS excitation energy, $\wcis$, in the doubles denominators defines the Hybrid partition (Hyb-CIS(2)) and reduces the bias to $-0.02$~eV. The substantial reduction in bias suggests that the doubles reference gap plays an important role in controlling the overcorrelation of the Canonical partition (Can-CIS(2)). The Hybrid partition uses the same doubles reference gap as CIS(D); the two methods differ in the Fock operator and in the treatment of the triples. This correspondence allows us to isolate how the CIS-state Fock operator and the treatment of the triples affect excitation energies relative to CIS(D).

Whether this gap remains appropriate with the CIS-state Fock operator is a separate question. Although Hyb-CIS(2) reduces the mean error relative to Can-CIS(2), appreciable errors remain for individual states and excitation classes. Its modification of the doubles reference gap amounts to adding $\delta_{\rm H}=\wzero-\wcis$ to each doubles denominator. We therefore regard the Hybrid partition as a particular state-dependent level shift and investigate whether varying the magnitude and form of the shift can further improve the accuracy.

Level shifts are widely used in perturbation theories to control contributions from small denominators and to modify the balance of correlation. Examples of the former are the real\cite{Roos1995LevelShift} and imaginary\cite{Forsberg1997ImagShift} shifts in CASPT2\cite{Andersson1990CASPT2,Park2020MRReview} and in spin-projected perturbation theory,\cite{Tsuchimochi2019SUPT2,Tsuchimochi2021SUPT2} and a real shift applied to excited-state M{\o}ller--Plesset perturbation theory (ESMP2)\cite{SheaNeuscamman2018ESMF,CluneSheaNeuscamman2020ESMP2} has been shown to improve excitation energies.\cite{Clune2023RegularizedESMP2} The ionization-potential--electron-affinity (IPEA) shift in CASPT2 addresses the imbalance between closed- and open-shell correlation.\cite{Ghigo2004IPEA,Zobel2017IPEA} Shifts of the excitation denominators have also been used to describe orbital relaxation in second-order theories.\cite{Mochizuki2008SelfEnergyShift,Mochizuki2009DysonPT2} Related approaches include regularized second-order methods,\cite{Stuck2013RegOOMP2,Lee2018RegularizedMP2,Shee2021kappaMP2,Battaglia2022RegularizedCASPT2} which damp contributions associated with small denominators, and spin-component scaling,\cite{Grimme2007SCSCISD,Grimme2012SCSReview} which rescales same-spin and opposite-spin pair contributions.

In this work, we pursue two objectives. First, we analyze CIS(2) in greater detail by deriving its connections to MP2 and CIS(D) and expanding the benchmark from the 100 states of our previous study to 334 states drawn from the QUEST datasets,\cite{Loos2018QUEST1,Loos2020QUEST3,Loos2021QUEST6,Veril2021QUEST} which extend earlier benchmark sets of vertical excitation energies.\cite{Schreiber2008Benchmark} After testing the CIS(D) connection with the HF Fock operator, we determine how using the CIS-state Fock operator changes the accuracy for valence, Rydberg, and charge-transfer excitations in the singlet and triplet benchmarks. We also compare the unshifted Canonical and Hybrid partitions to assess the role of the doubles reference gap and its relation to the composition of the CIS reference.

Second, we propose two variants of level shifts designed to preserve size-intensivity and investigate them as a means of improving both excitation energies and convergence of the amplitude equations in CIS(2). We test whether such shifts applied to Can-CIS(2) can improve upon unshifted Hyb-CIS(2) and whether additional shifts can further improve the Hybrid results.

This article is organized as follows. In Sec.~\ref{sec:theory}, we review CIS(2), derive its relations to MP2 and CIS(D), and construct the two size-intensive shifts. Section~\ref{sec:comp} describes the computational details. Section~\ref{sec:results} first tests the CIS(D) connection, then compares the unshifted partitions, analyzes how the CIS-state Fock operator changes the excitation energies of each class, and assesses the correction by the reference gap. We then quantify the contribution of the triples shift, the accuracy of the two shifts, the class and partition dependence of their parameters, and their effects on convergence. Conclusions are given in Sec.~\ref{sec:conclusions}.

\section{Theory}
\label{sec:theory}

\subsection{CIS(2)}
\label{sec:recap}

Here, we briefly review CIS(2) and define some quantities that will be required in the following sections. Throughout this paper, we use $i,j,k,l$ for occupied orbitals, $a,b,c,d$ for virtual orbitals, and $p,q,r,s$ for general orbitals, and Greek indices $\mu,\nu$ label distinct double substitutions with $a<b$ and $i<j$, or the corresponding contracted configurations. Unless noted otherwise, orbital indices denote spin orbitals.

The zeroth-order wave function is the CIS state,
\begin{align}
|S\rangle = \sum_{ai}c_{ai}|\Phi_i^a\rangle.
\end{align}
The zeroth-order Hamiltonian is constructed from the generalized Fock operator of the CIS state,
\begin{align}
f_{pq} = h_{pq} + G_{pq}[{\bf D}^S],
\label{eq:genfock}
\end{align}
where $D^S_{pq} = \langle S| a_q^\dagger a_p |S\rangle$ is an element of the CIS one-particle density matrix $\mathbf D^S$, and
\begin{align}
G_{pq}[{\bf D}^S] = \sum_{rs}\langle pr\Vert qs\rangle D^S_{sr}.
\end{align}
For singlet states, the occupied--virtual block is then projected orthogonal to the CIS coefficient vector to satisfy rigorous size-intensivity, as in Ref.~\onlinecite{Tsuchimochi2026paper1},
\begin{align}
 f_{ia}^{\perp}&=f_{ia}-\eta c_{ai},\\
 \eta&=\frac{\sum_{jb}f_{jb}c_{bj}}{\sum_{jb}c_{bj}^{2}}.
 \label{eq:fock_projection}
\end{align}
For the real orbitals used here, this gives $\sum_{ia}f_{ia}^{\perp}c_{ai}=0$. The removed component is proportional to $\langle\Phi_0|\hat F|S\rangle$, which couples the HF determinant to the CIS reference. For an excitation localized on fragment $A$, we refer to a noninteracting fragment $B$ that remains in its ground state as a ``spectator'' fragment. This matrix element also couples a double excitation confined to fragment $B$ to the corresponding contracted triple containing the CIS excitation on fragment $A$. Removing this component prevents such mixing and preserves the cancellation of the correlation contribution of fragment $B$ against its ground-state MP2 contribution. The occupied--occupied and virtual--virtual blocks are unchanged. For triplet states with an RHF reference, $c_{a\beta,i\beta}=-c_{a\alpha,i\alpha}$, whereas the Fock matrix is the same for both spins, so that the $\alpha$ and $\beta$ contributions to $\eta$ cancel, $\eta=0$, and $\mathbf f^{\perp}=\mathbf f$. The projection leaves the reference Fock expectation and the hole/particle block contractions unchanged.

Using the resulting Fock operator, we define the zeroth-order Hamiltonian in the same form as in ESMP2,\cite{SheaNeuscamman2018ESMF,CluneSheaNeuscamman2020ESMP2}

\begin{align}
\hat H_0 = \hat R(\Fock-\Ham)\hat R + \hat P\Ham\hat P + \hat Q\Fock\hat Q,
\label{eq:h0}
\end{align}
where $\Fock=\sum_{pq}f_{pq}^{\perp}a_p^\dagger a_q$ and $\hat R=|S\rangle\langle S|$. $\hat P$ projects onto the HF determinant and all its single substitutions, and $\hat Q=1-\hat P$. Since $|S\rangle$ is an eigenstate of $\hat H_0$ and the perturbation has no coupling to the other states in $\hat P$, the first-order correction lies in $\hat Q$.

Following the internally contracted formulation discussed in Ref.~\onlinecite{Tsuchimochi2026paper1}, we retain the complete HF-parent doubles space and the triples contracted with the CIS coefficients,
\begin{align}
|D_\mu\rangle &= \hat\tau_\mu|\Phi_0\rangle,\label{eq:Dmu}\\
|T_\mu\rangle &= \hat\tau_\mu|S\rangle,\label{eq:Tmu}
\end{align}
where $\hat\tau_\mu=a_a^\dagger a_b^\dagger a_j a_i$ is a particle--hole double excitation operator relative to the HF determinant $|\Phi_0\rangle$. The first-order wave function is then written as
\begin{align}
|\Psi^{(1)}\rangle
= \sum_\mu u_\mu|D_\mu\rangle
+ \sum_\mu t_\mu|T_\mu\rangle.
\label{eq:psi1}
\end{align}
This space contains the full Hamiltonian coupling $\hat Q\Ham|S\rangle$, while restricting the triples response to the internally contracted space. The contracted triples are not orthogonal and can be exactly or nearly linearly dependent, because different $\hat\tau_\mu$ acting on different components of $|S\rangle$ can generate the same determinant.\cite{Tsuchimochi2026paper1} Projecting the first-order equation
\begin{align}
(\hat H_0 - E^{(0)}_S)|\Psi^{(1)}\rangle = -(\Ham - \hat H_0)|S\rangle
\end{align}
onto the doubles and triples spaces, we obtain the coupled linear equations
\begin{align}
\begin{pmatrix} \mathbf{A}^{\mathrm{DD}} & \mathbf{A}^{\mathrm{DT}}\\ \mathbf{A}^{\mathrm{TD}} & \mathbf{A}^{\mathrm{TT}}\end{pmatrix}
\begin{pmatrix} \mathbf{u} \\ \mathbf{t} \end{pmatrix}
= -\begin{pmatrix} \mathbf{v}^{\mathrm{D}} \\ \mathbf{v}^{\mathrm{T}}\end{pmatrix},
\label{eq:linsys}
\end{align}
with $A^{\mathrm{XY}}_{\mu\nu} = \langle X_\mu|\Fock - E^{(0)}_S|Y_\nu\rangle$ and $E^{(0)}_S=\langle S|\Fock|S\rangle$, $v^{\mathrm D}_\mu = \langle D_\mu|\Ham|S\rangle$, and $v^{\mathrm T}_\mu = \langle T_\mu|\Ham|S\rangle$.

Once ${\bf u}$ and ${\bf t}$ are obtained, the CIS(2) excitation energy is given by
\begin{align}
\omega_{\mathrm{CIS(2)}} &= \wcis + E^{(2)}_S - E^{(2)}_0,
\label{eq:omega}\\
E^{(2)}_S &= \sum_{a<b,i<j}\bigl(u_{abij}v^{\mathrm{D}}_{abij}
 + t_{abij}v^{\mathrm{T}}_{abij}\bigr),
\label{eq:e2s}
\end{align}
where $E^{(2)}_S$ is the second-order energy correction to the CIS excited state $|S\rangle$, and $E^{(2)}_0$ is the MP2 correlation energy of the ground state. The explicit working equations are summarized in the Supporting Information.

The zeroth-order energy determines the reference gap subtracted in the amplitude equations. We denote the gaps used in the doubles and triples blocks by $\omega_{\rm D}$ and $\omega_{\rm T}$, respectively. The Canonical partition defined above uses $\omega_{\rm D}=\omega_{\rm T}=\wzero$, where $\wzero=E^{(0)}_S-\langle\Phi_0|\Fock|\Phi_0\rangle =\mathrm{Tr}[\mathbf f\,\bm\gamma]$ and $\bm\gamma=\mathbf D^S-\mathbf D^{\mathrm{HF}}$ is the CIS difference density. In our previous work,\cite{Tsuchimochi2026paper1} we also introduced two alternative partitions. The Block-diagonal partition uses $\omega_{\rm D}=\omega_{\rm T}=\wcis$, whereas the Hybrid partition uses $\omega_{\rm D}=\wcis$ and $\omega_{\rm T}=\wzero$. When $\wzero>\wcis$, replacing $\wzero$ by $\wcis$ increases the corresponding denominators.

The correlation of a noninteracting spectator fragment is carried by the contracted triples. Size-intensivity requires this contribution to recover the spectator MP2 correlation energy and cancel against the corresponding part of $E^{(2)}_0$. The conditions for this cancellation are discussed in Sec.~\ref{sec:sizeint}. We do not consider the Block-diagonal partition further because it is not size-intensive. Thus, $\omega_{\rm T}=\wzero$ throughout the following discussion.

\subsection{Relations to MP2 and CIS(D)}
\label{sec:lambda}
CIS(2) is consistent with its ground-state counterpart, MP2,\cite{Moller1934MP2} in the sense that its working equations naturally reduce to MP2 upon setting the explicit CIS coefficients to zero. In this limit, ${\bm \gamma} = {\bf 0}$, and the density and Fock matrices reduce to those of HF. It is then straightforward to show that the only nonzero elements in the first-order amplitude equations (\ref{eq:linsys}) are
\begin{align}
\left.\mathbf A^{\mathrm{DD}}\right|_{\mathbf c=\mathbf 0}&=\bm\Delta^{\mathrm{HF}} - \omega_{\mathrm D}\mathbf I,
\label{eq:mp2_doubles_block}\\
\left.\mathbf A^{\mathrm{TT}}\right|_{\mathbf c=\mathbf 0}&=\bm\Delta^{\mathrm{HF}},\\
\left.v^{\mathrm T}_{abij}\right|_{\mathbf c=\mathbf 0} &=w_{abij},
\label{eq:mp2_working_limit}
\end{align}
where $w_{abij}=\langle ab\Vert ij\rangle$ and $\bm\Delta^{\mathrm{HF}}$ is diagonal with elements $\Delta^{\mathrm{HF}}_{abij}=f^{\mathrm{HF}}_{aa}+f^{\mathrm{HF}}_{bb} -f^{\mathrm{HF}}_{ii}-f^{\mathrm{HF}}_{jj}$ in canonical HF orbitals, with $\mathbf f^{\rm HF}$ the HF Fock matrix. The remaining blocks $\mathbf A^{\mathrm{DT}}$, $\mathbf A^{\mathrm{TD}}$, and $\mathbf v^{\mathrm D}$ all vanish.

Equation~\eqref{eq:linsys} then admits $\mathbf u=\mathbf 0$, and the doubles channel makes no energy contribution. In contrast, the equation for the contracted triples yields
\begin{align}
t_{abij}&=t^{\mathrm{MP2}}_{abij}
=-\frac{w_{abij}}{\Delta^{\mathrm{HF}}_{abij}},\\
E^{(2)}_S&=-\sum_{a<b,i<j}
\frac{|w_{abij}|^2}{\Delta^{\mathrm{HF}}_{abij}},
\label{eq:mp2_energy_limit}
\end{align}
which is the MP2 correlation energy $E^{(2)}_0$.

We next examine the relation to CIS(D), retaining nonzero CIS coefficients and varying the density used to construct the Fock operator. To this end, we introduce an order parameter $\lambda$ that interpolates between the HF and CIS-state densities,
\begin{align}
\mathbf D(\lambda)&=\lambda\mathbf D^S+(1-\lambda)\mathbf D^{\rm HF}.
\label{eq:density_interpolation}
\end{align}
Thus, $\lambda=0$ uses the HF Fock operator, while $\lambda=1$ corresponds to standard CIS(2). 

At $\lambda=0$, the occupied--virtual Fock block vanishes by Brillouin's theorem. The D and T channels are therefore decoupled (see Eqs.~(S14) and (S15) in the Supporting Information), but their coupling vectors retain the CIS coefficients. In canonical HF orbitals, the doubles block is diagonal with elements $\Delta^{\mathrm{HF}}_{abij}-\omega_{\mathrm D}$, whereas the triples block retains the overlap and response of the CIS reference. We now compare the use of MP2 doubles amplitudes for the contracted triples in CIS(D) with the solution of this equation for the contracted triples.

The CIS(D) excitation energy can be written as\cite{HeadGordon1994CISD,Tsuchimochi2026paper1}
\begin{align}
\omega_{\mathrm{CIS(D)}} &= \wcis
 - \sum_{a<b,i<j}\frac{|v^{\mathrm D}_{abij}|^2}{\Delta^{\mathrm{HF}}_{abij}-\wcis}
\nonumber\\
&\quad + \sum_{a<b,i<j} t^{\mathrm{MP2}}_{abij}\,v^{\mathrm T}_{abij} - E^{(2)}_0,
\label{eq:cisd}
\end{align}
For the Hybrid choice $\omega_{\mathrm D}=\wcis$, the uncoupled doubles equation gives $u_{abij}=-v^{\mathrm D}_{abij}/(\Delta^{\mathrm{HF}}_{abij}-\wcis)$. Its energy contribution is identical to the second term of Eq.~\eqref{eq:cisd}, so the two energy expressions differ only in the triples amplitudes. Hyb-CIS(2) at $\lambda=0$ obtains them from $\mathbf A^{\mathrm{TT}}\mathbf t=-\mathbf v^{\mathrm T}$, whereas CIS(D) employs the MP2 amplitudes $\mathbf t^{\mathrm{MP2}}$ introduced above.

To understand this difference, we express $\mathbf v^{\rm T}$ and $\mathbf A^{\rm TT}$ in terms of HF quantities. First, we consider ${\bf v}^{\rm T}$. Let us write the CIS state as $|S\rangle =\hat C_1|\Phi_0\rangle $ with $\hat C_1=\sum_{ai}c_{ai}a_a^\dagger a_i$, and exploit the factorization $|T_\mu\rangle =\hat\tau_\mu\hat C_1|\Phi_0\rangle =\hat C_1|D_\mu\rangle $. Since $[\Ham,\hat C_1]$ is at most a two-body operator, $\langle T_\mu|[\hat H, \hat C_1] | \Phi_0\rangle = 0$, leading to the equality $v_\mu^{\rm T} \equiv \langle T_\mu|\hat C_1 \hat H| \Phi_0\rangle$. Introducing the resolution of the identity, we find
\begin{align}
\langle T_\mu|\hat C_1 \hat H| \Phi_0\rangle &= \langle T_\mu | \hat C_1 |\Phi_0\rangle \langle \Phi_0| \hat H | \Phi_0\rangle
\notag\\&+
\sum_{ai}\langle T_\mu | \hat C_1 |\Phi_i^a\rangle \langle \Phi_i^a| \hat H | \Phi_0\rangle
\notag\\&+
\sum_{a<b,i<j}\langle T_\mu | \hat C_1 |\Phi_{ij}^{ab}\rangle \langle \Phi_{ij}^{ab}| \hat H | \Phi_0\rangle
\\
&= \sum_\nu \langle T_\mu|T_\nu\rangle \langle D_\nu| \hat H |\Phi_0\rangle,
\end{align}
where the first two terms vanish in the first equality because the overlaps are zero. Therefore, by defining the overlap metric
\begin{align}
S_{\mu\nu} = \langle T_\mu| T_\nu\rangle,	
\end{align}
we have a simple relation
\begin{align}
\mathbf v^{\mathrm T}=\mathbf S\mathbf w.
\label{eq:vTMg}
\end{align}
Note that Eq.~(\ref{eq:vTMg}) also holds at $\lambda = 1$ since neither $\mathbf v^{\rm T}$ nor $\mathbf S$ involves the Fock operator.

Next, we consider $\mathbf A^{\rm TT}$ at $\lambda=0$. In canonical HF orbitals, a double excitation operator satisfies
\begin{align}
[\Fock,\hat\tau_\nu]
&=\Delta_\nu^{\mathrm{HF}}\hat\tau_\nu,
\end{align}
yielding
\begin{align}
A_{\mu\nu}^{\rm TT}
&=\langle T_\mu|(\Fock-E^{(0)}_S)\hat\tau_\nu|S\rangle
\notag\\
&=\langle T_\mu|[\Fock,\hat\tau_\nu]|S\rangle+\langle T_\mu|\hat\tau_\nu
                 (\Fock-E^{(0)}_S)|S\rangle
\notag\\
&=S_{\mu\nu}\Delta_\nu^{\rm HF}
 +\langle T_\mu|\hat\tau_\nu
                 (\Fock-E^{(0)}_S)|S\rangle.
\end{align}
The second term remains because the CIS reference is generally not an eigenstate of the HF Fock operator. Defining its residual as
\begin{align}
|\delta S\rangle&=(\Fock-E^{(0)}_S)|S\rangle,
\end{align}
we have $\langle S|\delta S\rangle=0$ and
\begin{align}
\mathbf A^{\rm TT}&=\mathbf S\bm\Delta^{\rm HF}+\mathbf B,
\label{eq:ATTsplit}\\
B_{\mu\nu}&=\langle T_\mu|\hat\tau_\nu|\delta S\rangle.
\label{eq:Bdef}
\end{align}

With Eqs.~(\ref{eq:vTMg}) and (\ref{eq:ATTsplit}), we finally arrive at the triples equation for $\lambda=0$ in the following different representation;
\begin{align}
\mathbf S\left(\bm\Delta^{\rm HF}\mathbf t+\mathbf w\right) + {\bf B}{\bf t}={\bf 0}.
\label{eq:MPres}
\end{align}
Hence, if $\mathbf B$ is omitted, the MP2 amplitudes satisfy the resulting equation, recovering CIS(D).

In summary, CIS(D) can be viewed as an approximation to Hyb-CIS(2) at $\lambda=0$ in which the residual Fock action on the CIS reference is neglected. If this residual has only a small effect on the triples energy, the two methods are expected to yield similar excitation energies. Section~\ref{sec:cisdlimit} tests this expectation, separating the effect of the triples response from that of the doubles reference gap.

\subsection{Size-intensive level shifts}
\label{sec:origin}
\label{sec:operator}
\label{sec:sizeint}

Given the relation between CIS(2) and CIS(D), we now consider how to correct the negative bias of the Canonical partition. The Hybrid partition replaces the doubles reference gap $\wzero$ by $\wcis$. Relative to Can-CIS(2), its change in the coupled amplitude matrix is
\begin{align}
\delta_{\rm H}&=\wzero-\wcis,\label{eq:hybrid_shift}\\
\mathbf A_{\rm Hybrid}-\mathbf A_{\rm Canonical}
&=\begin{pmatrix}\delta_{\rm H}\mathbf I^{\mathrm{DD}}&0\\0&0\end{pmatrix}.
\label{eq:hybrid_matrix}
\end{align}
For $\wcis<\wzero$, the Hybrid partition raises every doubles diagonal element by the same amount, leaving the triples and coupling blocks unchanged. This uniform doubles shift is expected to reduce the excessive correlation associated with the doubles response. Its effectiveness and state dependence are assessed in Sec.~\ref{sec:gapanalysis}.

We now generalize the magnitude and form of this correction through adjustable shifts of the amplitude equations, while retaining the Canonical reference gap. Extending the shift to the triples channel requires greater care to preserve size-intensivity. For an excitation localized on one fragment, the contracted triples include the product of that excitation and doubles on a noninteracting spectator fragment. Their correlation contribution cancels against the ground-state MP2 contribution of the spectator. A uniform shift $\varepsilon\mathbf I$ in the triples channel would change this contribution, but the cancellation could be retained by applying the same shift to the corresponding ground-state MP2 denominators. Here, however, we would rather retain the conventional MP2 ground-state correlation energy and seek to raise the CIS(2) excitation energies by modifying only the excited-state amplitude equations. Under this choice, the shift must leave the spectator MP2 contribution unchanged.

We therefore require the operator for the triples shift to satisfy two conditions. First, it must vanish for amplitudes whose four substitution indices belong entirely to the spectator fragment, thereby preserving its MP2 correlation contribution, and second, it must be positive semidefinite so that a positive shift parameter raises, rather than lowers, the denominators. To construct a simple operator satisfying both conditions, we use the CIS hole and particle density matrices,
\begin{align}
(\gamma_{\rm h})_{ij}&=\sum_a c_{ai}c_{aj},\\
(\gamma_{\rm p})_{ab}&=\sum_i c_{ai}c_{bi}.
\label{eq:gammap}
\end{align}
These matrices correspond to the detachment and attachment densities\cite{HeadGordon1995AttachDetach} and form the occupied and virtual blocks of the CIS difference density as $-\gh$ and $\gp$, respectively. Both are positive semidefinite and vanish on a noninteracting spectator fragment for a localized excitation. We let the hole density act on each occupied index and the particle density on each virtual index of the amplitude tensor. Adding these four actions defines the density-weighted operator $\Nhat$,
\begin{align}
(\Nhat\mathbf t)_{abij}
&=\sum_k\bigl[(\gamma_{\rm h})_{ik}t_{abkj}+(\gamma_{\rm h})_{jk}t_{abik}\bigr]\nonumber\\
&\quad +\sum_c\bigl[(\gamma_{\rm p})_{ac}t_{cbij}+(\gamma_{\rm p})_{bc}t_{acij}\bigr].
\label{eq:Nact}
\end{align}
Since the hole and particle density matrices are positive semidefinite, $\Nhat$ is also positive semidefinite. For a localized excitation, these densities vanish on a noninteracting spectator fragment, so $\Nhat$ annihilates amplitudes whose excitation indices all belong to that fragment. We note that this is not a unique choice, but perhaps the simplest one satisfying the two conditions.

The diagonal element of $\Nhat$ associated with a substitution $ij\to ab$ is the sum of its hole and particle density weights,
\begin{align}
N_{abij}&=(\gamma_{\rm h})_{ii}+(\gamma_{\rm h})_{jj}
 +(\gamma_{\rm p})_{aa}+(\gamma_{\rm p})_{bb}.
\label{eq:Nmu}
\end{align}
Thus, the diagonal part of the shift weights each substituted orbital by its participation in the CIS excitation. This weighting by occupation is reminiscent of the IPEA shift in CASPT2,\cite{Ghigo2004IPEA} although motivated here by the preservation of spectator correlation and positive semidefiniteness.

The doubles channel allows a more flexible choice, since double excitations confined to the spectator fragment make no contribution in this channel. Both a uniform shift $\varepsilon\mathbf u$ and a density-weighted shift $\varepsilon\Nhat\mathbf u$ preserve the zero amplitudes of doubles confined to that fragment. These choices give the two shift operators
\begin{align}
\mathbf L^{\mathrm I}&=\begin{pmatrix}\mathbf I^{\mathrm{DD}}&0\\0&{\bm{\mathcal N}}^{\mathrm{TT}}\end{pmatrix},
\label{eq:shiftIN}\\
\mathbf L^{\mathrm N}&=\begin{pmatrix}{\bm{\mathcal N}}^{\mathrm{DD}}&0\\0&{\bm{\mathcal N}}^{\mathrm{TT}}\end{pmatrix},
\label{eq:shiftNN}
\end{align}
where the superscript $X=\mathrm I,\mathrm N$ on $\mathbf L^{X}$ denotes the doubles operator; both choices use $\Nhat$ in the triples channel. Neither operator introduces couplings between noninteracting fragments, so both preserve the cancellation of the spectator fragment correlation contribution against its ground-state MP2 contribution, as in the unshifted theory\cite{Tsuchimochi2026paper1} (Table~S1 of the Supporting Information).

For $X=\mathrm I,\mathrm N$, the shifted Canonical matrix is
\begin{align}
\mathbf A^{X}(\varepsilon)&=\mathbf A+\varepsilon\mathbf L^{X}.
\label{eq:amplitude_shift}
\end{align}
We modify the amplitude equations directly and evaluate the excitation energy from Eq.~\eqref{eq:omega}.
A single shift parameter $\varepsilon$ is used on both channels and shared by all states within each prescription and is calibrated separately for the I- and N-shifts. The two prescriptions therefore share the triples shift at a given $\varepsilon$ and differ in the doubles shift and in the calibrated value of $\varepsilon$.

The relation to Hyb-CIS(2) is particularly transparent for the I-shift. Setting its parameter to the state-specific value $\delta_{\rm H}$ gives
\begin{align}
\mathbf A^{\mathrm I}(\delta_{\rm H})
&=\mathbf A_{\rm Hybrid}
 +\begin{pmatrix}0&0\\0&\delta_{\rm H}{\bm{\mathcal N}}^{\mathrm{TT}}\end{pmatrix}.
\label{eq:IN_hybrid_relation}
\end{align}
Thus the doubles blocks coincide, but the triples responses and excitation energies need not. 

The same operators can also be applied to Hyb-CIS(2). For example, relative to the unshifted Canonical matrix, the Hybrid matrix with an additional I-shift is
\begin{align}
\mathbf A_{\rm Hybrid}^{\mathrm I}(\varepsilon)-\mathbf A_{\rm Canonical}
&=\begin{pmatrix}(\delta_{\rm H}+\varepsilon)\mathbf I^{\mathrm{DD}}&0\\
0&\varepsilon{\bm{\mathcal N}}^{\mathrm{TT}}\end{pmatrix}.
\label{eq:hybrid_IN}
\end{align}
This prescription differs from shifted Can-CIS(2) and requires its own calibration. The primary accuracy comparison below uses Can-CIS(2); calculations with an additional shift in the Hybrid partition test the partition dependence.

Besides modifying excitation energies, the shifts are expected to improve convergence of the amplitude equations. As shown below, exact and near linear dependence among the contracted triples can produce zero and small eigenvalues even when the diagonal gaps are positive, hindering iterative solution when these directions are retained in the amplitude space. Since $\Nhat$ is positive semidefinite, the triples shift cannot lower any eigenvalue, and it lifts the redundant directions to the extent that they carry CIS hole or particle weight. 

Finally, we consider the computational cost of applying these level shifts. In the triples channel, the action of $\Nhat$ in Eq.~(\ref{eq:Nact}) is absorbed into the TT matrix--vector product, which already contains the contractions with the hole and particle densities of the CIS state, so the triples shift requires no additional contractions in either prescription. The two prescriptions therefore differ in cost only in the doubles channel. According to Eq.~(\ref{eq:Nact}), applying $\Nhat$ to the doubles amplitudes requires additional $O(o^3v^2)$ operations in the semicanonical basis, where the contraction with ${\bm \gamma}_{\rm p}$ is evaluated through the CIS coefficients of Eq.~(\ref{eq:gammap}). The I-shift therefore avoids the density-weighted contraction in the doubles channel.

\section{Computational details}
\label{sec:comp}

All calculations were performed using our in-house Python implementation of CIS(2) based on PySCF.~\cite{Sun2020PySCF} The core orbitals were frozen in the benchmark calculations, including CIS; the water/6-31G diagnostics in Sec.~\ref{sec:conv} correlate all electrons. The coupled equations were solved by right-preconditioned restarted GMRES~\cite{SaadSchultz1986GMRES} with the diagonal of the working matrix in the semicanonical basis, including the level shift when applied, as the preconditioner, until the relative residual $\|\mathbf A\mathbf x+\mathbf v\|/\|\mathbf v\|$ fell below $10^{-5}$, unless a tighter threshold is specified, where $\mathbf x=(\mathbf u,\mathbf t)$ and $\mathbf v=(\mathbf v^{\rm D},\mathbf v^{\rm T})$ are the vectors of Eq.~\eqref{eq:linsys}. GMRES was restarted every 20 iterations, except for the tests at the tighter threshold in Sec.~\ref{sec:conv}, which used a restart interval of 60 iterations.

The benchmark consists of 334 vertical excitations taken from the QUEST sets~\cite{Loos2018QUEST1,Loos2020QUEST3,Loos2021QUEST6,Veril2021QUEST}: 53 singlets and 47 triplets of QUEST\#1, 123 singlets and 88 triplets of QUEST\#3, and 23 intramolecular charge-transfer singlets of QUEST\#6. The aug-cc-pVTZ basis set was used for QUEST\#1 and \#3 and the cc-pVTZ basis set for QUEST\#6,~\cite{Dunning1989Basis,Kendall1992Aug,Woon1993Basis} and the results were compared with the published theoretical best estimates (TBEs). The charge-transfer (CT) class includes the hydrogen chloride state of QUEST\#1 in addition to the 23 QUEST\#6 states. States assigned as double excitations in the benchmark selection were excluded.

To obtain the shift dependence of the excitation energies and the optimal shift parameters, we calculated excitation energies at a series of $\varepsilon$ values for each partition and shift prescription on the same 334 states. The sampled values are listed in the Supporting Information. As shown below, the excitation energies depend approximately linearly on $\varepsilon$, and intermediate excitation energies were obtained accurately by monotone piecewise cubic Hermite interpolation (PCHIP) of the signed errors.\cite{Fritsch1980PCHIP,FritschButland1984} During this investigation, we identified discrepancies between our CIS(D) calculations and seven values published in the QUEST datasets. The corrected values used in this work are documented in Table~S2 of the Supporting Information.

\section{Results and discussion}
\label{sec:results}

\subsection{Recovery of CIS(D) at \texorpdfstring{$\lambda=0$}{lambda = 0}}
\label{sec:cisdlimit}

We first assess the numerical consequences of the relation derived in Sec.~\ref{sec:lambda}. While CIS(D) is recovered exactly when $\mathbf B$ is omitted from the triples equation of Hyb-CIS(2) at $\lambda=0$, the full equation retains $\mathbf B$ and need not give the same energy. We therefore examine how much this triples response changes the excitation energies. Table~\ref{tab:cisdlimit} compares the two for all 334 states of QUEST\#1, \#3, and \#6.

\begin{table}[tbp]
\caption{Comparison between Hyb-CIS(2) at $\lambda=0$ and CIS(D) in excitation energies. The mean and maximum absolute differences between the two methods are given in meV, and the MAEs relative to the TBE in eV.}
\label{tab:cisdlimit}
\begin{ruledtabular}
\begin{tabular}{lcccccc}
 & & \multicolumn{2}{c}{Difference (meV)} & & \multicolumn{2}{c}{MAE (eV)} \\
\cline{3-4}\cline{6-7}
Subset & $n$ & mean & max & & Hyb-CIS(2) & CIS(D) \\
\hline
QUEST\#1 singlets & 53 & 23 & 98 & & 0.28 & 0.28 \\
QUEST\#1 triplets & 47 & 10 & 27 & & 0.22 & 0.22 \\
QUEST\#3 singlets & 123 & 14 & 84 & & 0.20 & 0.21 \\
QUEST\#3 triplets & 88 & 9 & 36 & & 0.25 & 0.24 \\
QUEST\#6 singlets & 23 & 15 & 52 & & 0.38 & 0.40 \\
\hline
All & 334 & 14 & 98 & & 0.24 & 0.24 \\
\end{tabular}
\end{ruledtabular}
\end{table}

The two methods are indeed close numerically. The mean absolute difference is only $14$~meV, and 185 of the 334 states agree to within $10$~meV. The largest differences, up to $98$~meV, are found for valence singlets; for the ten largest, all of which are valence singlets of QUEST\#1 and QUEST\#3, the inclusion of $\mathbf B$ lowers the excitation energy. Both methods give a mean absolute error (MAE) of $0.24$~eV, and hence $\mathbf B$ has little effect on the overall accuracy. CIS(D) can therefore be regarded as a noniterative approximation to Hyb-CIS(2) at $\lambda=0$. The ground-state MP2 amplitudes are thus a judicious choice for the triples: they retain the accuracy of the full solution while requiring no iterative solution of the coupled amplitude equations. At $\lambda=1$, by contrast, the mean absolute difference from CIS(D) is $0.36$~eV for the same 334 states, so that the close correspondence with CIS(D) holds only when the HF Fock operator is used.

Given the close numerical correspondence between CIS(D) and Hyb-CIS(2) with the HF Fock operator ($\lambda=0$), we now turn to the accuracy of CIS(2) with the CIS-state Fock operator ($\lambda=1$), which is used in the benchmark calculations below.

\begin{table}[tbp]
\caption{Errors of CIS(2) and CIS(D). MAEs and MEs relative to the TBE are given in eV for the valence (V), Rydberg (R), and charge-transfer (CT) states. CIS(2) uses the CIS-state Fock operator ($\lambda=1$).}
\label{tab:unshifted}
\begin{ruledtabular}
\begin{tabular}{lrlrrr}
Subset & $n$ & & Can-CIS(2) & Hyb-CIS(2) & CIS(D)\\
\hline
V & 206 & MAE & 0.59 & 0.27 & 0.24\\
 & & ME & $-0.57$ & $-0.19$ & $+0.22$\\
R & 104 & MAE & 0.21 & 0.15 & 0.22\\
 & & ME & $-0.18$ & $+0.11$ & $-0.03$\\
CT & 24 & MAE & 0.54 & 0.28 & 0.39\\
 & & ME & $-0.49$ & $-0.03$ & $+0.32$\\
 \hline
 All & 334 & MAE & 0.47 & 0.24 & 0.24\\
 & & ME & $-0.44$ & $-0.09$ & $+0.15$\\
\end{tabular}
\end{ruledtabular}
\end{table}

\subsection{Accuracy of the Canonical and Hybrid partitions}
\label{sec:unshifted}

Before discussing the effect of level shifts, we compare Can-CIS(2) and Hyb-CIS(2) on the expanded benchmark of 334 states. Table~\ref{tab:unshifted} summarizes their errors alongside CIS(D). The Canonical MAE and ME are $0.47$ and $-0.44$~eV, whereas the Hybrid values are $0.24$ and $-0.09$~eV. These results confirm the tendency found in our previous study for the Canonical partition to substantially overestimate the correlation energy of excited states relative to the MP2 ground state. Note that, with the extended datasets, the Canonical MAE increases from $0.32$~eV for the 100 QUEST\#1 states,\cite{Tsuchimochi2026paper1} to $0.47$~eV, highlighting the importance of assessing the method across a broader range of molecules and excitation characters.

The improvement obtained with the Hybrid partition is achieved without changing the Fock operator or any block other than the doubles block. Thus, the doubles reference gap strongly affects the negative bias of Can-CIS(2) (see Eq.~(\ref{eq:hybrid_matrix})). Nevertheless, the improvement is not uniform across excitation classes. Hyb-CIS(2) reduces the valence and CT MAEs from $0.59$ and $0.54$~eV to $0.27$ and $0.28$~eV, respectively. For Rydberg states, the MAE decreases from $0.21$ to $0.15$~eV, but the ME changes sign, from $-0.18$ to $+0.11$~eV. Thus, the small overall Hybrid ME, $-0.09$~eV, results from the cancellation between the underestimation for valence states and the overestimation for Rydberg states. It is therefore important to analyze the errors separately for each excitation class, as we do in the following.

The comparison with CIS(D) also reproduces the main conclusion of our previous study.\cite{Tsuchimochi2026paper1} The CIS(D) MAEs are $0.24$, $0.22$, and $0.39$~eV for valence, Rydberg, and CT excitations, so that Hyb-CIS(2) is slightly less accurate for valence excitations but more accurate for Rydberg excitations and, as the expanded benchmark reveals, for CT excitations. The advantage over CIS(D) thus extends to both Rydberg and CT states, which involve pronounced redistribution of the excited electron density.

In the next section, we investigate the origin of this dependence of the accuracy of CIS(2) on the excitation character.

\subsection{Effect of the CIS-state Fock operator}
\label{sec:fockeffect}

Since Hyb-CIS(2) at $\lambda=0$ closely reproduces CIS(D) (Table~\ref{tab:cisdlimit}), the differences between CIS(D) and the Hybrid results at $\lambda=1$ in Table~\ref{tab:unshifted} mainly reflect the replacement of the HF density by the CIS-state density in the Fock operator. This replacement leaves the overall MAE nearly unchanged, at $0.24$~eV, but shifts the excitation energies in different directions among the classes. The ME changes from $+0.22$ to $-0.19$~eV for valence states and from $+0.32$ to $-0.03$~eV for CT states, but from $-0.03$ to $+0.11$~eV for Rydberg states. For valence states, this substantial downward shift changes the MAE only slightly, from $0.24$ to $0.27$~eV.

For Rydberg states, the reduction in MAE reflects the state dependence of the energy changes rather than their positive mean. These changes are anticorrelated with the errors of Hyb-CIS(2) at $\lambda=0$ ($r=-0.80$), and the standard deviation of the errors decreases from $0.31$ to $0.19$~eV upon using the CIS-state Fock operator.

\begin{table}[tbp]
\caption{Hole and particle contributions to the Fock-density effect in Hyb-CIS(2). The excitation energies change when $\mathbf G$ of the hole density ($-\gh$), the particle density ($\gp$), or both ($\bm\gamma$) is added to the HF Fock operator; the mean changes (eV) are listed for the DD/TT blocks, the DT/TD couplings, and all blocks (Total), together with the mean norm (eV) of the projected occupied--virtual block $\mathbf G_{\rm vo}^{\perp}$ of each density.}
\label{tab:fockblocks}
\begin{ruledtabular}
\begin{tabular}{llrrrr}
Class & Density & DD/TT & DT/TD & Total & $\|\mathbf G_{\rm vo}^{\perp}\|$\\
\hline
V & Hole & $-0.23$ & $+0.08$ & $-0.16$ & 6.9\\
 & Particle & $-0.32$ & $-0.03$ & $-0.37$ & 5.8\\
 & Both & $-0.52$ & $+0.10$ & $-0.40$ & 3.9\\
\hline
R & Hole & $+0.03$ & $+0.33$ & $+0.34$ & 7.1\\
 & Particle & $-0.25$ & $-0.04$ & $-0.30$ & 1.4\\
 & Both & $-0.16$ & $+0.30$ & $+0.15$ & 6.3\\
\hline
CT & Hole & $-0.24$ & $+0.12$ & $-0.12$ & 6.2\\
 & Particle & $-0.28$ & $-0.03$ & $-0.33$ & 4.7\\
 & Both & $-0.50$ & $+0.15$ & $-0.33$ & 3.9\\
\end{tabular}
\end{ruledtabular}
\end{table}

\begin{figure}[tbp]
\includegraphics[width=\columnwidth]{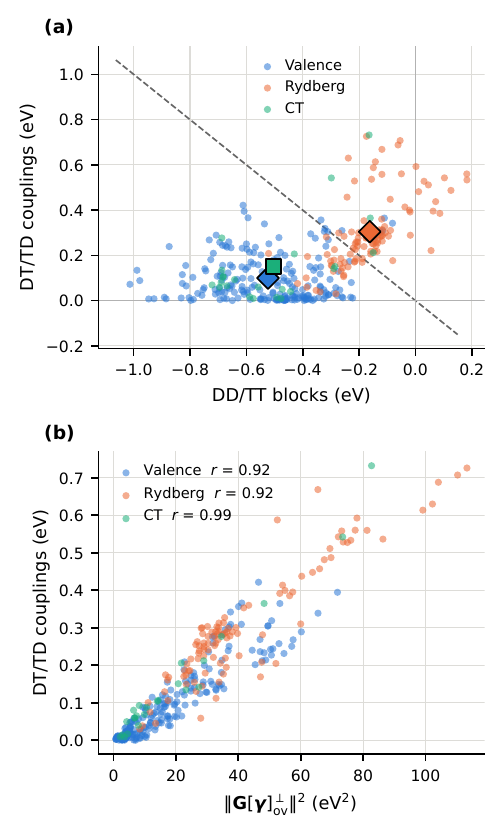}
\caption{Block decomposition of the Fock-density effect in Hyb-CIS(2). The excitation energies change from the HF to the CIS-state Fock operator. (a) The DD/TT contribution is plotted against the DT/TD contribution; large symbols are class means, and the dashed line indicates complete cancellation. (b) The DT/TD contribution is plotted against $\|\mathbf G[\bm\gamma]_{\rm vo}^{\perp}\|^2$, with the correlation coefficient $r$ of each class.}
\label{fig:fockblocks}
\end{figure}

To clarify how this replacement shifts the excitation energies differently among the classes, we decompose the resulting change of the Fock operator. Recall that replacing the HF density by the CIS-state density changes the Fock operator by $\mathbf G[\bm\gamma]$ of Eq.~(\ref{eq:genfock}). Accordingly, we separate $\mathbf G[\bm\gamma]$ into its occupied--occupied, virtual--virtual, and occupied--virtual blocks. The occupied--occupied and virtual--virtual blocks modify the DD/TT blocks of the amplitude equations, whereas the projected occupied--virtual block, $\mathbf G[\bm\gamma]_{\rm vo}^{\perp}$ of Eq.~(\ref{eq:fock_projection}), modifies the DT/TD couplings (see the Supporting Information). We solve the Hyb-CIS(2) equations with each part added separately to the HF Fock operator. Since $\mathbf G$ is linear in the density, the difference between the potentials of the particle and hole densities is given by $\mathbf G[\bm\gamma]=\mathbf G[\gp]-\mathbf G[\gh]$ (Sec.~\ref{sec:operator}); we therefore also add $\mathbf G[\gp]$ and $\mathbf G[-\gh]$ separately to obtain their individual contributions.

The results are summarized in Table~\ref{tab:fockblocks} and Figure~\ref{fig:fockblocks}. For the full difference density ${\bm\gamma}$, the separately evaluated DD/TT and DT/TD contributions are nearly additive; $\mathbf G[\bm\gamma]$ produces competing effects through the DD/TT blocks and the DT/TD couplings. Changing the DD/TT blocks lowers the excitation energies on average in all three classes, whereas changing the DT/TD couplings raises them in all states. For Rydberg states, the former lowering is smaller and the latter increase is larger than for valence states, resulting in a net upward shift. For valence and CT states, the lowering through the DD/TT blocks dominates. Figure~\ref{fig:fockblocks}(a) shows this competition for individual states. For most Rydberg states, the increase through the DT/TD couplings outweighs the lowering through the DD/TT blocks, whereas the latter dominates for most valence states. The four CT states above the cancellation line, the two states of twisted DMABN and one state of each dipeptide, follow the Rydberg pattern rather than that of the other CT states.

To understand why their balance differs among excitation classes, we consider the DD/TT and DT/TD contributions separately. The smaller lowering through the DD/TT blocks for Rydberg states arises mainly from the hole density. As shown in Table~\ref{tab:fockblocks}, the DD/TT blocks of the hole density alone lower the valence and CT excitation energies by $0.23$ and $0.24$~eV, respectively, but leave the Rydberg excitation energies nearly unchanged ($+0.03$~eV), whereas those of the particle density lower all three classes by similar amounts, $0.25$--$0.32$~eV.

Table~\ref{tab:fockblocks} also shows that the larger increase through the DT/TD couplings for Rydberg states is associated with a larger $\mathbf G[\bm\gamma]_{\rm vo}^{\perp}$. Within each class, the DT/TD contribution increases nearly linearly with $\|\mathbf G[\bm\gamma]_{\rm vo}^{\perp}\|^2$, as shown in Figure~\ref{fig:fockblocks}(b). In $\mathbf G[\bm\gamma]_{\rm vo}^{\perp}=\mathbf G[\gp]_{\rm vo}^{\perp}-\mathbf G[\gh]_{\rm vo}^{\perp}$, the particle block partially cancels the hole block in every state (Table~S3 of the Supporting Information). For Rydberg states, $\mathbf G[\gp]_{\rm vo}^{\perp}$ is much smaller, leaving a larger $\mathbf G[\bm\gamma]_{\rm vo}^{\perp}$ than in the valence and CT classes (Table~\ref{tab:fockblocks}). This small particle contribution is consistent with the diffuse particle density, whose Coulomb potential varies slowly over the region of the occupied orbitals.

Now, it is worth noting that $\mathbf G[\bm\gamma]_{\rm vo}^{\perp}$ is related to orbital relaxation of the excited state.\cite{Gilbert2008MOM,Hait2021OODFT,SheaNeuscamman2018ESMF} This is because, at HF orbitals, the occupied--virtual orbital gradient of the CIS energy, $g_{ai}=\partial E_{\rm CIS}/\partial\kappa_{ai}$ with $\kappa_{ai}$ the parameters of the occupied--virtual orbital rotations, which drives this relaxation, is given by
\begin{align}
\mathbf g_{\rm vo}&=2\left(\mathbf G[\bm\gamma]_{\rm vo}+\bm\tau\right),\label{eq:cisgrad}\\
\bm\tau&=\mathbf G[\mathbf c^{\mathsf T}]_{\rm vv}\,\mathbf c-\mathbf c\,\mathbf G[\mathbf c^{\mathsf T}]_{\rm oo},\label{eq:cisgrad_t}
\end{align}
where $\mathbf c$ is the virtual--occupied matrix of the CIS coefficients $c_{ai}$ and $\mathbf G[\mathbf c^{\mathsf T}]$ is the potential of the transition density (see Eq.~(S22) of the Supporting Information for the general expression). The first term arises from the difference density and the second from the transition density. Since $\bm\tau$ is small compared with $\mathbf G[\bm\gamma]_{\rm vo}$ for Rydberg and CT states, $\mathbf G[\bm\gamma]_{\rm vo}^{\perp}$ is closely aligned with the CIS orbital gradient for these states (Table~S3 and Figure~S1a), and their norms are nearly proportional, $\|\mathbf g_{\rm vo}\|\simeq 2\|\mathbf G[\bm\gamma]_{\rm vo}^{\perp}\|$. The larger $\|\mathbf G[\bm\gamma]_{\rm vo}^{\perp}\|$ of the Rydberg states in Table~\ref{tab:fockblocks} therefore reflects a larger orbital gradient, that is, a stronger driving force for orbital relaxation. Accordingly, the DT/TD contribution correlates strongly with $\|\mathbf g_{\rm vo}\|$ within each class (Figure~S1b). Since the orbital gradient vanishes at orbitals optimized for the excited state and the Fock contribution dominates this gradient for Rydberg and CT states at HF orbitals, orbital optimization is expected to reduce the occupied--virtual Fock coupling and its upward contribution to the CIS(2) excitation energies.

In summary, we have shown that replacing the HF Fock operator by the CIS-state Fock operator in Hyb-CIS(2) produces two competing effects. Their balance, which depends on how the hole density changes the DD/TT blocks and on how strongly the particle block cancels the hole block in $\mathbf G[\bm\gamma]_{\rm vo}^{\perp}$, accounts for the different directions of the average changes in the excitation energies among the three classes. Combined with the errors at $\lambda=0$, these average changes account for the changes in the ME described above, so that the Hybrid errors differ in sign between valence and Rydberg states.

\subsection{Doubles reference gap and reference composition}
\label{sec:gapanalysis}
\begin{figure}[tbp]
\includegraphics[width=\columnwidth]{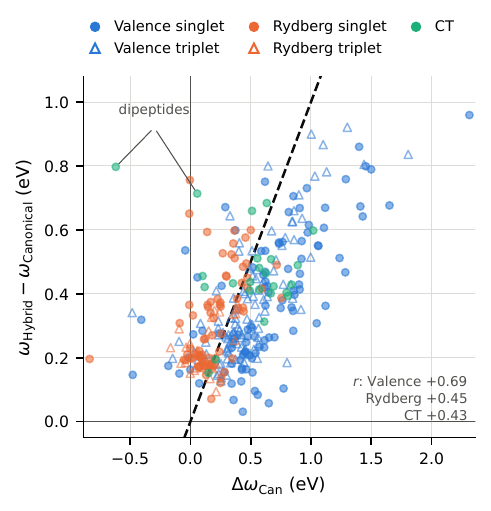}
\caption{Correction supplied by the Hybrid reference gap. $\omega_{\rm Hybrid}-\omega_{\rm Canonical}$ is plotted against $\Delta\omega_{\rm Can}=\omega_{\rm TBE}-\omega_{\rm Canonical}$, with the correlation coefficient $r$ of each class; the two labeled dipeptide states are excluded from the CT value of $r$. The dashed line indicates $\omega_{\rm Hybrid}=\omega_{\rm TBE}$.}
\label{fig:hyb}
\end{figure}

We now turn from the Fock operator to the correction supplied by the Hybrid doubles reference gap. Figure~\ref{fig:hyb} plots $\omega_{\rm Hybrid}-\omega_{\rm Canonical}$ against $\Delta\omega_{\rm Can}=\omega_{\rm TBE}-\omega_{\rm Canonical}$, which is positive when Can-CIS(2) underestimates the excitation energy and is the correction it requires. Hyb-CIS(2) raises all excitation energies and thereby reduces the negative bias. For most Rydberg states it raises the energy by more than $\Delta\omega_{\rm Can}$, whereas for most valence states it falls short, increasingly so for the states with the largest errors. The Hybrid correction therefore reduces the bias but does not match $\Delta\omega_{\rm Can}$ uniformly across states.

To characterize this state dependence, we next ask whether the unshifted errors are associated with the natural transition orbital (NTO) composition of the CIS reference.\cite{Martin2003NTO} For this purpose, we use the effective rank of the CIS coefficient matrix, expressed by the hole--electron entanglement measure of Plasser,\cite{Plasser2012Exciton,Plasser2016Entanglement}
\begin{align}
Z_{\mathrm{HE}} &= 2^{S_{\mathrm{HE}}},
\label{eq:zhe}\\
S_{\mathrm{HE}} &= -\sum_k \sigma_k^2\log_2 \sigma_k^2,
\end{align}
where $\sigma_k$ are the singular values of the $\alpha$ block of $\mathbf c$ renormalized to unity, for both singlet and $M_S=0$ triplet states, so that $\sigma_k^2$ is the weight of the $k$th NTO pair and $\sum_k\sigma_k^2=1$. $Z_{\mathrm{HE}}$ represents the effective number of contributing NTO pairs, which is unity for an excitation with a single NTO pair and equal to $m$ for $m$ equally weighted pairs.

\begin{figure}[tbp]
\includegraphics[width=\columnwidth]{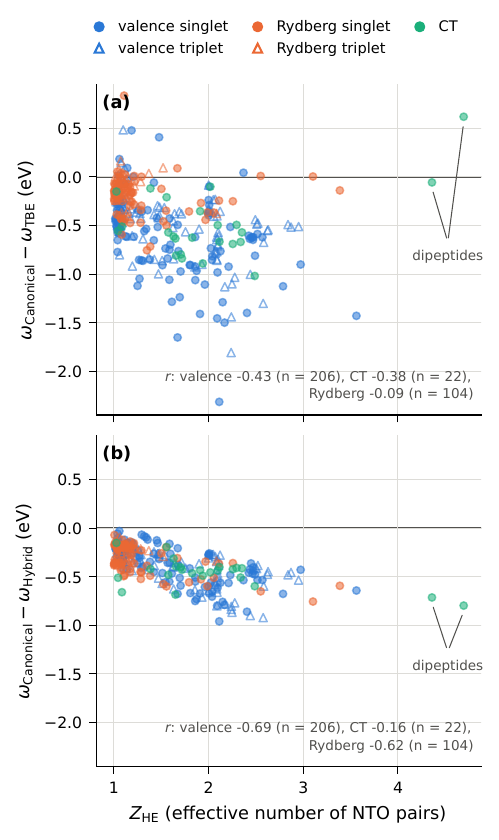}
\caption{Canonical errors and Hybrid corrections against the effective number of NTO pairs. (a) The error of Can-CIS(2), $\omega_{\rm Canonical}-\omega_{\rm TBE}$, and (b) the correction supplied by the Hybrid reference gap, $\omega_{\rm Canonical}-\omega_{\rm Hybrid}$, are plotted against $Z_{\mathrm{HE}}$ on the same scale. The correlation coefficients of each class are given in each panel; the two dipeptide states with $Z_{\mathrm{HE}}>4$ are excluded from the CT values.}
\label{fig:zhe}
\end{figure}

Figure~\ref{fig:zhe}(a) shows the error of Can-CIS(2) against $Z_{\mathrm{HE}}$. For valence states, the error becomes more negative as $Z_{\mathrm{HE}}$ increases, and most of the CT states follow the same trend, whereas no clear correlation is observed for Rydberg states, most of which have $Z_{\mathrm{HE}}$ near unity. Figure~\ref{fig:zhe}(b) shows the correction supplied by the Hybrid reference gap, $\omega_{\rm Canonical}-\omega_{\rm Hybrid}$, on the same scale. For the valence states, the correction grows with $Z_{\mathrm{HE}}$ ($r=-0.69$) in parallel with the Canonical error, and this dependence also holds within molecules ($r=-0.61$ after subtracting the molecular means of $Z_{\mathrm{HE}}$ and of the correction). The Hybrid errors of the valence states therefore depend only weakly on $Z_{\mathrm{HE}}$ ($r=-0.11$, compared with $-0.43$ for the Canonical errors). Part of this dependence of the correction comes from $\delta_{\rm H}$ of Eq.~(\ref{eq:hybrid_shift}), which increases with $Z_{\mathrm{HE}}$ ($r=+0.68$ for valence and $+0.56$ for Rydberg states). For the Rydberg states, the correction also grows with $Z_{\mathrm{HE}}$ although the Canonical errors are already small, so that Hyb-CIS(2) overestimates the few Rydberg states with large $Z_{\mathrm{HE}}$. For the CT states, the correction is nearly independent of $Z_{\mathrm{HE}}$, and the Hybrid errors retain the Canonical trend. The Hybrid reference gap thus removes most of the dependence of the valence errors on the reference composition, but overcorrects the Rydberg states with large $Z_{\mathrm{HE}}$ and leaves the trend of the CT states unchanged.

\begin{figure}[tbp]
\includegraphics[width=\columnwidth]{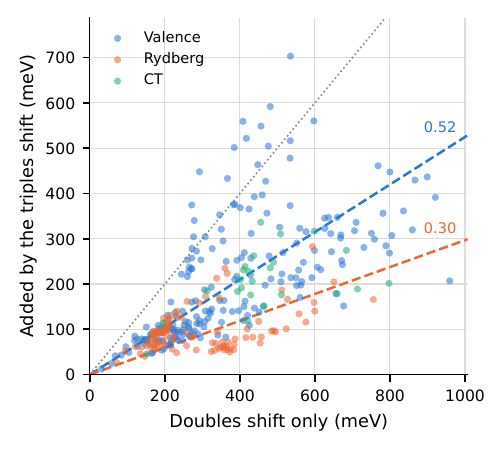}
\caption{Doubles and triples parts of the shift. The change of the excitation energy at $\varepsilon=\delta_{\rm H}$ is split into the doubles part (Hyb-CIS(2) minus unshifted Can-CIS(2)) and the triples part (Can-CIS(2) with the I-shift minus Hyb-CIS(2)). The dotted line indicates equal parts, and the dashed lines are fits through the origin for the valence and Rydberg states (slopes $0.52$ and $0.30$, $r=0.64$ and $0.45$).}
\label{fig:tsplit}
\end{figure}

\subsection{Energy contribution of the triples shift}
\label{sec:Tshift_energy}
Since the proposed shifts change not only the DD block but also the TT block, it is important to isolate the contribution of the latter and determine its magnitude. Equation~\eqref{eq:IN_hybrid_relation} provides a direct test for this purpose. Here, we evaluate the I-shift excitation energies at $\varepsilon=\delta_{\rm H}$ and compare them with Hyb-CIS(2), which has the same doubles block, so that the difference gives the contribution of the triples shift by itself. The distribution of $\delta_{\rm H}$ is given in Figure~S2 of the Supporting Information. Figure~\ref{fig:tsplit} plots the contributions of the doubles and triples shifts, the former obtained as $\omega_{\rm Hybrid}-\omega_{\rm Canonical}$ and the latter as $\omega^{\rm I}_{\rm Canonical}(\delta_{\rm H})-\omega_{\rm Hybrid}$. On average, the doubles shift raises the excitation energies by 357~meV, and the triples shift raises them by a further 173~meV, about one third of the mean total change. For several valence states, the doubles and triples contributions are comparable, whereas the triples contribution is generally smaller relative to the doubles contribution for Rydberg states.

\begin{figure}[tbp]
\includegraphics[width=\columnwidth]{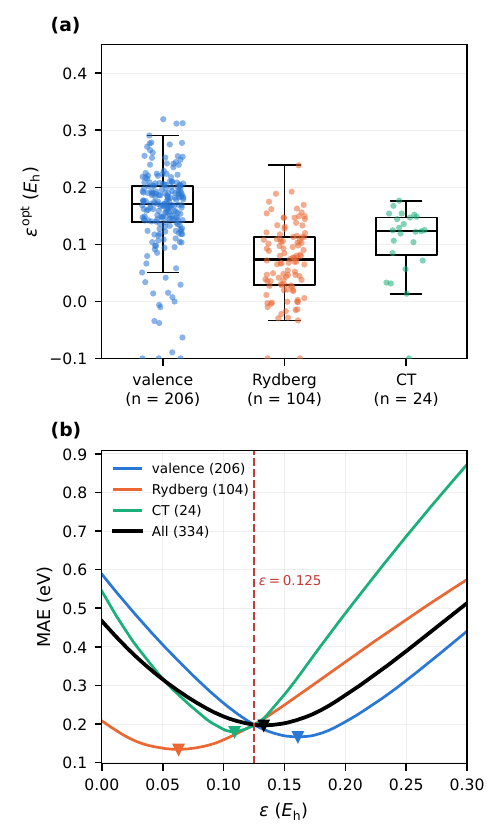}
\caption{Optimal parameters and class MAEs of Can-CIS(2) with the I-shift. (a) The state-specific parameter $\varepsilon^{\rm opt}\simeq\Delta\omega_{\rm Can}/\alpha$, estimated from a linear fit of each excitation energy against $\varepsilon$, is shown for each class; boxes give the quartiles and medians, and values outside the plotted range are placed at its edges. (b) The MAE of each class and of all states is plotted against $\varepsilon$, with the excitation energies interpolated between the computed parameters by PCHIP, and the minima are shown by triangles. The dashed line in (b) indicates the common parameter $0.125~E_{\rm h}$.}
\label{fig:lin}
\end{figure}

\begin{table*}[tbp]
\caption{Errors of shifted Can-CIS(2) and Hyb-CIS(2). MAEs and MEs are in eV; Can-CIS(2) uses $\varepsilon^{\mathrm I}=0.125$ and $\varepsilon^{\mathrm N}=0.160~E_{\rm h}$, and Hyb-CIS(2) uses $\varepsilon^{\mathrm I}=0.030$ and $\varepsilon^{\mathrm N}=0.035~E_{\rm h}$. }
\label{tab:main}
\begin{ruledtabular}
\begin{tabular}{lrlrrrrrrr}
 & & & \multicolumn{3}{c}{Can-CIS(2)} & \multicolumn{3}{c}{Hyb-CIS(2)} & \\
\cline{4-6}\cline{7-9}
Subset & $n$ & & Unshifted & I & N & Unshifted & I & N & CIS(D)\\
\hline
V & 206 & MAE & 0.59 & 0.20 & 0.22 & 0.27 & 0.20 & 0.21 & 0.24\\
 & & ME & $-0.57$ & $-0.09$ & $-0.10$ & $-0.19$ & $-0.09$ & $-0.09$ & $+0.22$\\
R & 104 & MAE & 0.21 & 0.20 & 0.22 & 0.15 & 0.21 & 0.21 & 0.22\\
 & & ME & $-0.18$ & $+0.18$ & $+0.21$ & $+0.11$ & $+0.19$ & $+0.19$ & $-0.03$\\
CT & 24 & MAE & 0.54 & 0.20 & 0.18 & 0.28 & 0.25 & 0.25 & 0.39\\
 & & ME & $-0.49$ & $+0.16$ & $+0.11$ & $-0.03$ & $+0.12$ & $+0.10$ & $+0.32$\\
\hline
All & 334 & MAE & 0.47 & 0.20 & 0.22 & 0.24 & 0.21 & 0.21 & 0.24\\
 & & ME & $-0.44$ & $+0.01$ & $+0.01$ & $-0.09$ & $+0.02$ & $+0.01$ & $+0.15$\\
\end{tabular}
\end{ruledtabular}
\end{table*}

\subsection{Accuracy of uniform and density-weighted doubles shifts}
\label{sec:accuracy}
To assess whether shifts applied to Can-CIS(2) can overcome the limitations of the Hybrid reference gap found in Sec.~\ref{sec:gapanalysis}, we compare the I- and N-shifts of Sec.~\ref{sec:operator}. Before calibrating the parameters, we examine how the excitation energies respond to the shift. The Canonical excitation energies depend nearly linearly on $\varepsilon$ for either shift, $\omega(\varepsilon)\simeq\omega_{\rm Canonical}+\alpha\,\varepsilon$, with a state-dependent slope $\alpha$ (see Figure~S3), so that the shift parameter that reproduces the TBE of each state is $\varepsilon^{\rm opt}\simeq\Delta\omega_{\rm Can}/\alpha$. This linear estimate also applies to states whose $\varepsilon^{\rm opt}$ lies outside the computed range of $\varepsilon$ ($0\le\varepsilon\le0.30~E_{\rm h}$), where it is obtained by extrapolation. The linear estimates of $\varepsilon^{\rm opt}$ for the I-shift for different classes are summarized in Figure~\ref{fig:lin}(a), showing that valence states require larger parameters than Rydberg states. Accordingly, we evaluated the MAE of the I-shift Can-CIS(2) excitation energies relative to the TBE for each class at each $\varepsilon$, using the excitation energies interpolated by PCHIP, and plotted the results in Figure~\ref{fig:lin}(b). 

As expected, the valence MAE is minimized at a larger parameter than the Rydberg MAE, so that a common parameter must compromise between the two classes. The CT MAE varies most steeply with $\varepsilon$, because the CT excitation energies respond most strongly to the shift, whereas the valence and Rydberg MAEs are less sensitive to the choice of the parameter. We therefore determine the common parameter by minimizing the larger of the valence and Rydberg MAEs of QUEST\#1 and \#3, with the CT states excluded so that they serve as an external test. This gives $\varepsilon^{\mathrm I}=0.125~E_{\rm h}$, shown as the dashed line in Figure~\ref{fig:lin}(b); the CT curve also passes through the crossing of the valence and Rydberg curves at nearly this value. The same criterion gives $\varepsilon^{\mathrm N}=0.160~E_{\rm h}$ for the N-shift.

Table~\ref{tab:main} compares Can-CIS(2) and Hyb-CIS(2) with the I-shift and the N-shift. Relative to unshifted Hyb-CIS(2), Can-CIS(2) with the I-shift lowers the MAE from $0.24$ to $0.20$~eV overall, from $0.27$ to $0.20$~eV for valence, and from $0.28$ to $0.20$~eV for CT states, but raises the Rydberg MAE from $0.15$ to $0.20$~eV, with the ME increasing from $+0.11$ to $+0.18$~eV. The overcorrection in Rydberg found for the Hybrid partition in Sec.~\ref{sec:gapanalysis} thus persists with a common parameter. Again, a shift with a common parameter changes the distribution of accuracy among the classes, and a compromise between them remains.

Comparing the I- and N-shifts in Can-CIS(2), we find that the I-shift gives slightly lower MAEs for the valence and Rydberg states, $0.20$~eV compared with $0.22$~eV for both classes. On the other hand, for CT, the N-shift gives a lower MAE of $0.18$~eV compared with $0.20$~eV for the I-shift. At their calibrated parameters, the two prescriptions thus give different balances of accuracy among the excitation classes. The MAEs and the optimal parameters for each benchmark set are given in Tables~S4 and S5 of the Supporting Information.

Hyb-CIS(2) provides a further comparison in Table~\ref{tab:main}. The same fitting criterion as above gives optima of $\varepsilon^{\mathrm I}=0.030$ and $\varepsilon^{\mathrm N}=0.035~E_{\rm h}$. Direct calculations at these parameters give an overall MAE of $0.21$~eV for both prescriptions, compared with $0.24$~eV without an additional shift and $0.20$~eV for Can-CIS(2) with the I-shift. The dependence of the common parameters on the fitting criterion and on the subset of states is examined in Sec.~S8 of the Supporting Information.

For the present benchmark, the I-shift is the preferable prescription with a common parameter in both partitions, under the fitting criterion above and also in the overall MAE, although the differences are small for Hyb-CIS(2). The CT advantage of the N-shift in Can-CIS(2) raises the question whether it persists when the parameter is calibrated for each class.

\subsection{Class dependence of the shift parameter}
\label{sec:optimal}
Since the MAEs of the valence and Rydberg states are minimized at different parameters [Figure~\ref{fig:lin}(b)], a common parameter cannot be optimal for both classes and gives a compromise between them.

Table~\ref{tab:opt} quantifies this compromise by minimizing the MAE separately for each excitation class, shift prescription, and partition. For Can-CIS(2), both prescriptions require a substantially larger parameter for valence than for Rydberg excitations. This ordering follows Figure~\ref{fig:hyb}, in which Hybrid raises most valence states by less than $\Delta\omega_{\rm Can}$ and most Rydberg states by more; for Hyb-CIS(2), the optimal nonnegative parameter is zero for both prescriptions in the Rydberg class. The valence advantage of the I-shift persists after separate calibration, with MAEs of $0.17$ and $0.20$~eV for the I- and N-shifts. For Rydberg states the two prescriptions become comparable. The CT advantage of the N-shift also nearly disappears, from $0.02$~eV at the common parameters to below $0.01$~eV at the optimal parameter of each class. Much of the CT difference under a common parameter therefore reflects the compromise required to describe all classes with one value.

\begin{table}[tb]
\caption{Optimal shift parameters $\varepsilon^\star$ ($E_{\rm h}$) for each excitation class. The parameters minimize the MAE (eV) of each class and are calibrated separately for each prescription and partition.}
\label{tab:opt}
\begin{ruledtabular}
\begin{tabular}{lrrrr}
 & \multicolumn{2}{c}{I} & \multicolumn{2}{c}{N}\\
Class & $\varepsilon^\star$ & MAE & $\varepsilon^\star$ & MAE\\
\hline
\multicolumn{5}{c}{Can-CIS(2)}\\
V & 0.160 & 0.17 & 0.205 & 0.20 \\
R & 0.065 & 0.13 & 0.065 & 0.14 \\
CT & 0.110 & 0.18 & 0.150 & 0.18 \\
\hline
\multicolumn{5}{c}{Hyb-CIS(2)}\\
V & 0.065 & 0.18 & 0.080 & 0.18 \\
R & 0.000 & 0.15 & 0.000 & 0.15 \\
CT & 0.030 & 0.25 & 0.045 & 0.24 \\
\end{tabular}
\end{ruledtabular}
\end{table}

The residual errors also reveal limitations of a nonnegative shift. For states whose excitation energies are already overestimated without a shift, a positive shift raises the energy further and thus always worsens the error. The largest residual errors are documented in the Supporting Information.

\subsection{Effect of the level shift on convergence}
\label{sec:conv}

As discussed in Sec.~\ref{sec:operator}, the level shift also raises the eigenvalues of the redundant and nearly redundant directions of the contracted triples, which can hinder the iterative solution in both partitions. We examine the convergence of the amplitude equations with the diagonal preconditioner in GMRES, using the I-shift and reporting the N-shift for comparison. We first confirm this mechanism for a small system and then assess the convergence for the 334 benchmark states and at a tighter threshold.

\begin{figure}[tbp]
\includegraphics[width=\columnwidth]{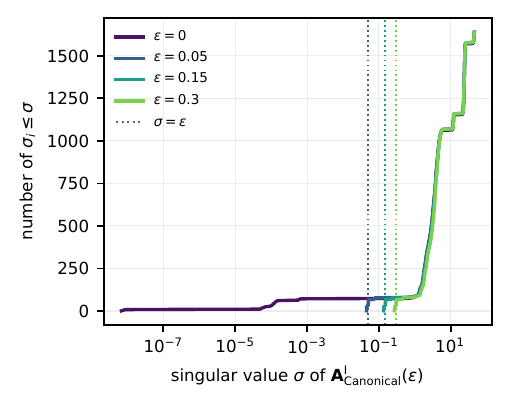}
\caption{Singular values of the working matrix. $\mathbf A^{\rm I}_{\rm Canonical}(\varepsilon)$ is that of the $S_1$ state of water (6-31G) in the pair-symmetric representative coordinates. Dotted lines are drawn at $\sigma=\varepsilon$.}
\label{fig:band}
\end{figure}

Figure~\ref{fig:band} shows the singular values of the working matrix $\mathbf A^{\rm I}_{\rm Canonical}(\varepsilon)$ for the S$_1$ state of water (6-31G) with different $\varepsilon$ values. Without the shift, the exact and near linear dependencies of the contracted triples produce a band of singular values close to zero, and the shift lifts this band so that the smallest singular value lies close to $\varepsilon$ (dotted lines).

For this state, the unshifted working matrix has 14 eigenvalues below $10^{-8}~E_{\rm h}$, which correspond to exact and nearly exact linear dependencies of the contracted triples, and no negative eigenvalues. The restriction of $\bm{\mathcal N}$ to these directions has eigenvalues between $1.0$ and $2.0$, so that the shift lifts all of them, and at $\varepsilon=0.15~E_{\rm h}$ the smallest eigenvalue of the shifted matrix equals $\varepsilon$ within numerical precision. The shifted system is therefore nonsingular.

Table~\ref{tab:conv} lists the numbers of GMRES iterations required to reach a relative residual of $10^{-5}$ for the 334 states, without the shift and with the I- and N-shifts in both partitions. All states converge in both partitions with and without the shift.

\begin{table}[tbp]
\caption{Convergence of the amplitude equations. The numbers of GMRES iterations per state are given at a relative residual of $10^{-5}$, with the common parameters of Table~\ref{tab:main}.}
\label{tab:conv}
\begin{ruledtabular}
\begin{tabular}{lllccc}
method & shift & $\varepsilon$ & mean & median & max \\
\hline
Can-CIS(2) & Unshifted & 0 & 24.8 & 20.5 & 63 \\
Can-CIS(2) & I & 0.125 & 12.2 & 11 & 29 \\
Can-CIS(2) & N & 0.160 & 11.6 & 10 & 27 \\
Hyb-CIS(2) & Unshifted & 0 & 24.7 & 21 & 63 \\
Hyb-CIS(2) & I & 0.030 & 15.9 & 14 & 38 \\
Hyb-CIS(2) & N & 0.035 & 15.5 & 14 & 37 \\
\end{tabular}
\end{ruledtabular}
\end{table}

For Can-CIS(2), the I-shift approximately halves the mean number of iterations, while for Hyb-CIS(2) it reduces the mean by about one third. The N-shift gives similar convergence in both partitions. This similarity is consistent with the density-weighted triples shift, common to both prescriptions, playing the dominant role in improving convergence.

\begin{figure}[tbp]
\includegraphics[width=\columnwidth]{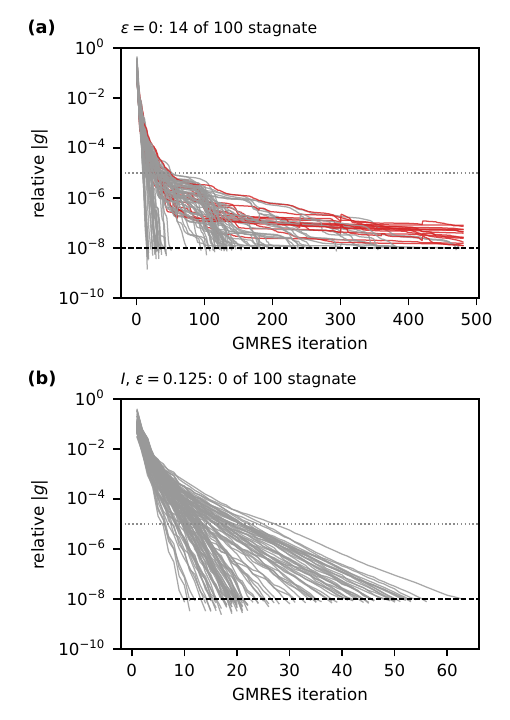}
\caption{Convergence histories of Can-CIS(2). The relative residual is plotted per iteration for the 100 QUEST\#1 states (a) without the shift and (b) with the I-shift at $\varepsilon=0.125~E_{\rm h}$, with a target of $10^{-8}$. Red lines are the states that do not reach the target, 14 in (a) and none in (b). The dotted line indicates the default threshold of $10^{-5}$.}
\label{fig:conv}
\end{figure}

The effect of the shift becomes more evident at a tighter threshold. Figure~\ref{fig:conv} presents the convergence behavior of Can-CIS(2) without and with the I-shift at $\varepsilon=0.125~E_{\rm h}$ for the 100 QUEST\#1 states, with a target relative residual of $10^{-8}$. Without the shift, as shown in Figure~\ref{fig:conv}(a), 14 states fail to converge within 480 iterations, with their residuals stagnating above the threshold (red curves). Their excitation energies, however, change by less than $0.1$~meV after the residual falls below $10^{-5}$. With the shift, on the other hand, all 100 states converge within 63 iterations (Figure~\ref{fig:conv}(b)). The shift therefore enables the amplitude equations to meet the tighter residual threshold even when the excitation energies have already stabilized during the iterations, which may be important for analytical gradients.

\section{Conclusions}
\label{sec:conclusions}
To better understand CIS(2) and improve its performance, we have clarified its connections to MP2 and CIS(D), assessed its accuracy on an expanded benchmark, and developed two size-intensive level shifts of its amplitude equations. Under the formal substitution $\mathbf c=0$, the working equations reduce to MP2 for both the Canonical and Hybrid partitions, independently of the order parameter $\lambda$. They also preserve the ground-state MP2 correlation on a noninteracting spectator fragment, making the method size-intensive\cite{Tsuchimochi2026paper1}. For Hyb-CIS(2) at $\lambda=0$, neglecting the residual Fock action on the CIS reference recovers CIS(D). Including the additional triples response changed the excitation energies only slightly.

Unshifted Hyb-CIS(2) with the CIS-state Fock operator was slightly less accurate than CIS(D) for valence excitations but more accurate for Rydberg excitations, consistent with our previous study. The expanded benchmark further revealed improved accuracy for CT excitations. Since Hyb-CIS(2) with the HF Fock operator closely reproduces CIS(D), these differences mainly reflect the use of the CIS-state Fock operator. To understand how this change of the Fock operator affects the excitation energies, we analyzed the contributions from the individual blocks of the amplitude equations. We found that changes in the DD/TT blocks lower the excitation energies on average, whereas changes in the DT/TD couplings raise them. For Rydberg states, the lowering is smaller and the increase is larger. The larger coupling contribution is associated with weaker cancellation between the particle and hole contributions to the occupied--virtual Fock block. The DT/TD contribution also correlated strongly with the norm of the CIS orbital gradient, suggesting that orbital optimization for the excited state may reduce this upward contribution. By comparing Canonical and Hybrid with the same Fock operator, we also showed the strong effect of the doubles reference gap on the negative bias. For valence states, replacing this gap substantially weakened the correlation between the error and the effective number of NTO pairs, although the correction remained uneven across states and excitation classes.

To preserve the ground-state correlation contribution of noninteracting fragments, we developed a density-weighted level shift for the triples channel using the CIS hole and particle densities. The doubles channel permits either an identity shift or a density-weighted shift. We found that the triples shift also made a substantial contribution to the change in the excitation energies, showing that the effect of the shift cannot be attributed to the doubles shift alone.

With a calibrated common parameter, Can-CIS(2) with the I-shift improved upon unshifted Hyb-CIS(2) in overall accuracy and for valence and CT excitations, but worsened the Rydberg accuracy. Our benchmark calculations also showed that, for Can-CIS(2), the I-shift gave better overall accuracy and lower valence and Rydberg MAEs than the N-shift at their independently calibrated common parameters. Density weighting gave a lower MAE for CT states, but this advantage largely disappeared when the parameters were calibrated separately for each class. Thus, although both doubles operators preserve size-intensivity, the present accuracy comparison favored the uniform choice. Hyb-CIS(2) also benefited from an additional shift, with a smaller parameter than Can-CIS(2).

The shifts also substantially reduced the number of iterations in both partitions, and the I-shift removed the stagnation of the residual observed for Can-CIS(2) at a tight convergence threshold. The similar convergence behavior of the two prescriptions is consistent with their common triples shift playing an important role. Neither prescription increases the formal scaling of the amplitude solve, while the uniform doubles shift avoids an additional density-weighted contraction.

Although a common parameter improved the overall accuracy, the optimal parameters differed between valence and Rydberg excitations for both prescriptions. This limitation motivates the development of state-dependent parameters, which are expected to provide more accurate excitation energies, together with an assessment of how their performance depends on the shift operator and partition.
\begin{acknowledgments}
This work was supported by the JST FOREST Program, Grant No. JPMJFR223U. Generative artificial intelligence (Claude by Anthropic) was used to organize the benchmark data, generate the data-processing scripts, and improve the readability and language of the manuscript.
\end{acknowledgments}

\section*{Associated Content}
\subsection*{Supporting Information}
Working equations in semicanonical spin orbitals; numerical verification of size-intensivity; corrected reference values and state assignment; the occupied--virtual block of $\mathbf G[\bm\gamma]$ and the CIS orbital gradient; distribution of $\delta_{\rm H}$; shift grids and parameter search; dependence on the shift parameter; accuracy of the I- and N-shifts for each benchmark set, optimal parameters for each set and class, and parameter stability; and the largest remaining errors (PDF). Excitation energies, shift grids, and analyses for each state, together with the GMRES histories and the singular values of Figure~\ref{fig:band} (XLSX).

\section*{Data and Software Availability}
The data underlying this study are available in the published article and its Supporting Information, which provides the corrected reference values, the excitation energies and shift grids of every benchmark state, and the decompositions of the Fock operator. The molecular geometries are those of the QUEST database. The calculations were performed using an in-house implementation built on PySCF,~\cite{Sun2020PySCF} which also contains developments associated with ongoing, unpublished work and is therefore not currently distributed. The complete working equations of CIS(2), from which it can be implemented independently, are provided in Sec.~S1 of the Supporting Information and in that of Ref.~\citenum{Tsuchimochi2026paper1}, and the level shift is defined in full in Sec.~\ref{sec:operator}.

\bibliography{refs}

\clearpage
\onecolumngrid
\setcounter{section}{0}\setcounter{equation}{0}\setcounter{figure}{0}\setcounter{table}{0}
\begin{center}
{\large\bfseries Supporting Information}
\end{center}
\renewcommand{\theequation}{S\arabic{equation}} \renewcommand{\thesection}{S\arabic{section}} \renewcommand{\thetable}{S\arabic{table}} \renewcommand{\thefigure}{S\arabic{figure}} 

\section{Working equations in semicanonical spin orbitals}
\label{si:working}
The spin-orbital working equations derived in Ref.~\onlinecite{Tsuchimochi2026paper1} are collected here in the semicanonical representation used in the calculations. The occupied indices are $i,j,k,l$, virtual indices are $a,b,c,d$, and general spin-orbital indices are $p,q,r,s$. The amplitudes are antisymmetric under exchange of either pair of occupied or virtual indices. Greek indices $\mu,\nu$ label distinct double substitutions with $a<b$ and $i<j$.

We use real orbitals and coefficients. The occupied and virtual blocks of the CIS-state generalized Fock matrix are diagonal, with elements $f_{ii}$ and $f_{aa}$; the occupied--virtual blocks need not vanish. Define
\begin{align}
\Delta_{ai}&=f_{aa}-f_{ii},\\
\Delta_{abij}&=\Delta_{ai}+\Delta_{bj},\\
\tilde{\Delta}_{ai}&=\Delta_{ai}c_{ai},\\
\gamma_{ki}&=-\sum_c c_{ci}c_{ck},\\
\gamma_{ac}&=\sum_k c_{ak}c_{ck},\\
\tilde h_{ki}&=\sum_c \tilde{\Delta}_{ck}c_{ci},\\
\eta&=\sum_{kc}f_{kc}c_{ck}.
\end{align}
Here all indices are spin orbitals, $\sum_{ai}c_{ai}^2=1$, and $\wzero=\sum_{ai}\tilde{\Delta}_{ai}c_{ai}$. With this normalization, $\eta$ is the parameter of Eq.~(5) of the main text, and the Fock projection, which removes the CIS component from the occupied--virtual block, gives $\eta=0$. The trace terms are retained below to display the unprojected form. The triples reference gap is $\omega_{\rm T}=\wzero$ throughout, with $\omega_{\rm D}=\wzero$ for Can-CIS(2) and $\omega_{\rm D}=\wcis$ for Hyb-CIS(2). We use $\mathcal P(ab)x_{abij}=x_{abij}-x_{baij}$ and similarly for $\mathcal P(ij)$.

Define the additional intermediates
\begin{align}
(\mathbf t\mathbf f)_{bj}&=\sum_{ai}f_{ia}t_{abij},\\
(\mathbf u\mathbf c)_{bj}&=\sum_{ai}c_{ai}u_{abij},\\
(\mathbf t\mathbf c)_{bj}&=\sum_{ai}c_{ai}t_{abij},\\
(\tilde{\bm\Delta}\mathbf t)_{bj}&=\sum_{ck}\tilde{\Delta}_{ck}t_{bcjk},\\
X_{abij}&=c_{ai}\bigl[(\Delta_{ai}+\Delta_{bj}-\wzero)
 (\mathbf t\mathbf c)_{bj}+(\tilde{\bm\Delta}\mathbf t)_{bj}\bigr].
\end{align}
Writing $\bm{\sigma}^{\rm XY}[\mathbf x]=\mathbf A^{\rm XY}\mathbf x$, all four blocks are
\begin{align}
\sigma^{\rm DD}[\mathbf u]_{abij}&=(\Delta_{abij}-\omega_{\rm D})u_{abij},\\
\sigma^{\rm DT}[\mathbf t]_{abij}&=
 \mathcal P(ab)\mathcal P(ij)c_{ai}(\mathbf t\mathbf f)_{bj}-\mathcal P(ab)\sum_k c_{ak}\left(\sum_c f_{kc}t_{cbij}\right)\nonumber\\
&\quad-\mathcal P(ij)\sum_k\left(\sum_c f_{kc}c_{ci}\right)t_{abkj}+\eta t_{abij},\\
\sigma^{\rm TD}[\mathbf u]_{abij}&=
 \mathcal P(ab)\mathcal P(ij)f_{ai}(\mathbf u\mathbf c)_{bj}-\mathcal P(ab)\sum_k f_{ka}\left(\sum_c c_{ck}u_{cbij}\right)\nonumber\\
&\quad-\mathcal P(ij)\sum_k\left(\sum_c f_{ic}c_{ck}\right)u_{abkj}+\eta u_{abij},\\
\sigma^{\rm TT}[\mathbf t]_{abij}&=
 \Delta_{abij}t_{abij}
 +\mathcal P(ab)\mathcal P(ij)X_{abij}\nonumber\\
&\quad+\mathcal P(ij)\sum_k\bigl[(\Delta_{abij}-\wzero)\gamma_{ki}-\tilde h_{ki}\bigr]t_{abkj}\nonumber\\
&\quad-\mathcal P(ab)\sum_k c_{ak}\left[\sum_c\bigl((\Delta_{abij}-\wzero)c_{ck}+\tilde{\Delta}_{ck}\bigr)t_{cbij}\right].
\label{eq:si_working_sigma}
\end{align}
These expressions give $\bm{\sigma}^{\rm DD}[\mathbf u]+\bm{\sigma}^{\rm DT}[\mathbf t]=-\mathbf v^{\rm D}$ and $\bm{\sigma}^{\rm TD}[\mathbf u]+\bm{\sigma}^{\rm TT}[\mathbf t]=-\mathbf v^{\rm T}$.

The right-hand sides are
\begin{align}
v^{\mathrm{D}}_{abij} &= \mathcal P(ij)\sum_c\langle ab||cj\rangle c_{ci}
 - \mathcal P(ab)\sum_k\langle kb||ij\rangle c_{ak}
 + \mathcal P(ij)\mathcal P(ab)\,f^{\rm HF}_{ai}c_{bj},
\label{eq:si_working_vd}\\
v^{\mathrm{T}}_{abij} &= \langle ab||ij\rangle 
 + \mathcal P(ij)\mathcal P(ab)\sum_{ck}\langle bc||jk\rangle c_{ai}c_{ck}
\nonumber\\
&\quad + \mathcal P(ij)\sum_k\gamma_{ki}\langle ab||kj\rangle 
 - \mathcal P(ab)\sum_c\gamma_{ca}\langle cb||ij\rangle.
\label{eq:si_working_vt}
\end{align}
Here $\langle pq\Vert rs\rangle$ denotes an antisymmetrized spin-orbital two-electron integral. The matrix $f^{\rm HF}$ in Eq.~\eqref{eq:si_working_vd} is the HF Fock operator expressed in the same orbitals as the other tensors; it is distinct from the generalized Fock matrix $f$. The bare-integral term in Eq.~\eqref{eq:si_working_vt} uses the normalization $\sum_{ai}c_{ai}^2=1$, as does the leading $\Delta_{abij}t_{abij}$ term of the triples action. Formally setting the explicit coefficients $c_{ai}$ to zero in these working equations removes the doubles coupling vector and the doubles--triples couplings. The resulting equation for $t_{abij}$ gives the MP2 doubles amplitudes, as discussed in the main text.

For a level shift, the coupled equations become
\begin{align}
\bm\sigma^{\rm DD}[\mathbf u]+\bm\sigma^{\rm DT}[\mathbf t]
 +\varepsilon\mathbf L^{\mathrm{DD}}\mathbf u&=-\mathbf v^{\rm D},\\
\bm\sigma^{\rm TD}[\mathbf u]+\bm\sigma^{\rm TT}[\mathbf t]
 +\varepsilon\Nhat^{\mathrm{TT}}\mathbf t&=-\mathbf v^{\rm T}.
\end{align}
The doubles operator is $\mathbf L^{\mathrm{DD}}=\mathbf I^{\mathrm{DD}}$ for the I-shift and $\mathbf L^{\mathrm{DD}}=\Nhat^{\mathrm{DD}}$ for the N-shift. With $\gamma_{\rm h}=-\gamma_{\rm oo}$ and $\gamma_{\rm p}=\gamma_{\rm vv}$, the density-weighted action on either amplitude tensor $\mathbf x$ is
\begin{align}
(\Nhat\mathbf x)_{abij}
&=\sum_k\bigl[(\gamma_{\rm h})_{ik}x_{abkj}+(\gamma_{\rm h})_{jk}x_{abik}\bigr]\nonumber\\
&\quad +\sum_c\bigl[(\gamma_{\rm p})_{ac}x_{cbij}+(\gamma_{\rm p})_{bc}x_{acij}\bigr].
\end{align}
The shift leaves the right-hand sides and the doubles--triples couplings unchanged.

\section{Numerical verification of size-intensivity}
\label{si:sinum}
Table~\ref{tab:si_sizeint} lists the change of the excitation energy when a helium atom is added to water.

\begin{table}[h]
\caption{Size-intensivity test. Change (eV) of the excitation energy of the S$_1$ state of water when a helium atom is placed at a distance $R$, relative to isolated water (8.571490~eV for the I-shift and 8.634340~eV for the N-shift). Can-CIS(2)/6-31G with the full shift operators $\mathbf L^{\mathrm I}$ ($\varepsilon=0.125~E_{\rm h}$) and $\mathbf L^{\mathrm N}$ ($\varepsilon=0.160~E_{\rm h}$); amplitude equations converged to a relative residual of $10^{-9}$.}
\label{tab:si_sizeint}
\begin{ruledtabular}
\begin{tabular}{lrr}
$R$ (\AA) & I & N\\
\hline
10 & $5.9\times10^{-13}$ & $5.6\times10^{-13}$\\
20 & $6.4\times10^{-13}$ & $6.7\times10^{-13}$\\
30 & $4.7\times10^{-13}$ & $4.7\times10^{-13}$\\
\end{tabular}
\end{ruledtabular}
\end{table}

\section{Reference values and state assignment}
\label{si:refdata}
Each benchmark state is assigned to the CIS root of the same irreducible representation whose CIS(D) energy is closest to the published CIS(D) value; ambiguous cases are resolved by orbital character. A published CIS(D) value is regarded as erroneous when its deviation from our value far exceeds the agreement obtained for the other states of the same molecule. The seven corrected values are listed in Table~\ref{tab:si_errata}.

\begin{table}[h]
\caption{Corrected CIS(D) reference values. Published CIS(D)/aug-cc-pVTZ values (eV) of QUEST\#3\cite{Loos2020QUEST3} and, for formaldehyde, QUEST\#1\cite{Loos2018QUEST1} are compared with the values of this work. ``Duplicate'' marks a published value that repeats an adjacent table entry; for these, scans of 60 (pyrazine) and 80 (tetrazine) roots find no root of the same symmetry near the published value.}
\label{tab:si_errata}
\begin{ruledtabular}
\begin{tabular}{llrrl}
Molecule & State & Published & This work & Error\\
\hline
Pyrazine & $1\,^1B_{1g}$ & 7.31 & 6.441 & duplicate\\
Cyclopropenone & $1\,^1B_1$ & 4.27 & 4.368 & value\\
Pyrrole & $1\,^1B_2$ & 6.62 & 6.737 & value\\
Tetrazine & $2\,^3A_u$ & 5.74 & 4.958 & duplicate\\
Tetrazine & $1\,^3B_{1g}$ & 4.46 & 4.565 & value\\
Cyclopropenone & $1\,^3A_2$ & 5.96 & 5.913 & value\\
Formaldehyde & $2\,^3A_1$ & 7.68 & 7.577 & value\\
\end{tabular}
\end{ruledtabular}
\end{table}

\section{Occupied--virtual block of \texorpdfstring{$\mathbf G[\gamma]$}{G[gamma]}}
\label{si:fov}
Let $\mathbf c$ denote the CIS coefficients arranged as a virtual--occupied matrix, $\mathbf f^{\rm HF}$ the Fock matrix of the HF determinant built from the current orbitals, and $\mathbf f=\mathbf f^{\rm HF}+\mathbf G[\bm\gamma]$ the generalized Fock matrix of the CIS state [Eq.~(2) of the main text]. Here, $\bm\gamma_{\rm h}$ and $\bm\gamma_{\rm p}$ are those of Eq.~(34) of the main text, and $\kappa_{ai}$ are the parameters of the occupied--virtual orbital rotations. The occupied--virtual orbital gradient of the CIS energy, $g_{ai}=\partial E/\partial\kappa_{ai}$, is
{\begin{align}
\mathbf g_{\rm vo}&=2\left[\mathbf f_{\rm vo}-\left(\bm\gamma_{\rm p}\mathbf f^{\rm HF}_{\rm vo}+\mathbf f^{\rm HF}_{\rm vo}\bm\gamma_{\rm h}\right)+\bm\tau\right],
\label{eq:si_grad}
\end{align}}
with $\bm\tau$ of Eq.~(43) of the main text. At HF orbitals, $\mathbf f^{\rm HF}_{\rm vo}=\mathbf 0$ gives Eq.~(42) of the main text. At orbitals optimized for the excited state, as in ESMF,\cite{SheaNeuscamman2018ESMF} $\mathbf g_{\rm vo}=\mathbf 0$ gives
{\begin{align}
\mathbf f_{\rm vo}&=\left(\bm\gamma_{\rm p}\mathbf f^{\rm HF}_{\rm vo}+\mathbf f^{\rm HF}_{\rm vo}\bm\gamma_{\rm h}\right)-\bm\tau,
\label{eq:si_stationary}
\end{align}}
so that $\mathbf f_{\rm vo}$ need not vanish.

\begin{table}[htbp]
\caption{Alignment of the occupied--virtual blocks. Mean cosine between $\mathbf G[\gh]_{\rm vo}^{\perp}$ and $\mathbf G[\gp]_{\rm vo}^{\perp}$ and median cosine between $\mathbf G[\bm\gamma]_{\rm vo}^{\perp}$ and the CIS orbital gradient at HF orbitals for the valence, Rydberg, and CT states.}
\label{tab:si_fov}
\begin{ruledtabular}
\begin{tabular}{lrr}
Class & $\cos(\mathbf G[\gh]_{\rm vo}^{\perp},\mathbf G[\gp]_{\rm vo}^{\perp})$ & $\cos(\mathbf g_{\rm vo},\mathbf G[\bm\gamma]_{\rm vo}^{\perp})$\\
\hline
V & $+0.82$ & 0.914\\
R & $+0.63$ & 0.995\\
CT & $+0.75$ & 0.959\\
\end{tabular}
\end{ruledtabular}
\end{table}

\begin{figure}[htbp]
\includegraphics[width=\textwidth]{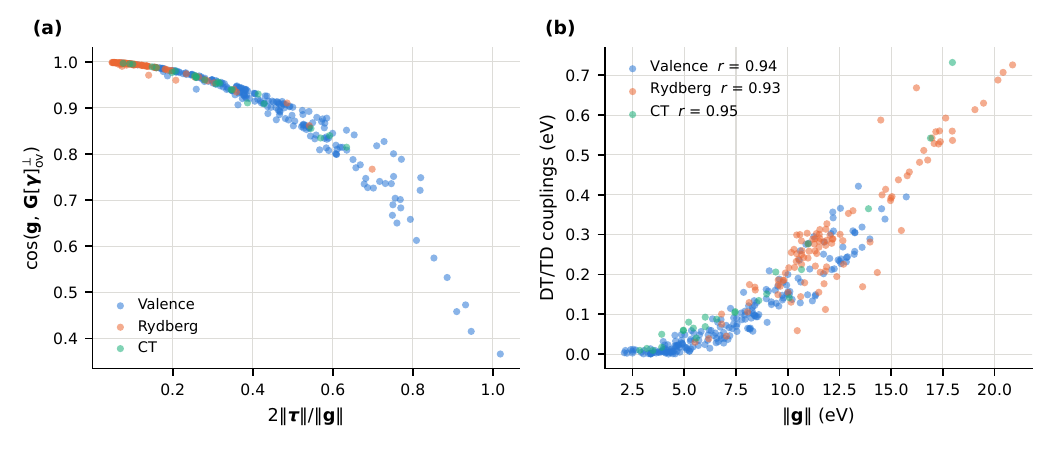}
\caption{Occupied--virtual block and the CIS orbital gradient at HF orbitals. (a) Cosine between $\mathbf g_{\rm vo}$ and $\mathbf G[\bm\gamma]_{\rm vo}^{\perp}$ against the fraction $2\|\bm\tau\|/\|\mathbf g_{\rm vo}\|$ of the transition-density term. (b) Contribution of the DT/TD couplings to the change from the HF to the CIS-state Fock operator against $\|\mathbf g_{\rm vo}\|$, with Pearson correlation coefficients within each class.}
\label{fig:si_gradient}
\end{figure}

\section{Hybrid doubles-denominator increments}
\label{si:deltaH}

For all benchmark states, we evaluate $\delta_{\rm H}=\omega^{(0)}-\omega_{\rm CIS}$ using the CIS-state generalized Fock operator and the same CIS states as in the energy calculations. The increment is positive for every state, with a median of $0.134~E_{\rm h}$.

\begin{figure}[h]
\includegraphics[width=\textwidth]{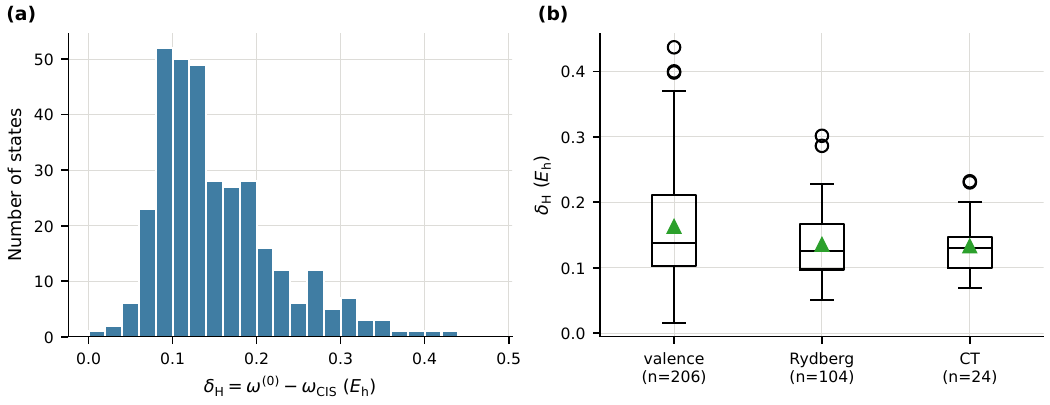}
\caption{Hybrid doubles-denominator increments. Distribution of $\delta_{\rm H}$ (a) for all states and (b) for each excitation class. Boxes show the interquartile range, horizontal lines the medians, and triangles the means. Whiskers extend to the most distant points within 1.5 interquartile ranges of each box, and points beyond them are shown as circles.}
\end{figure}

\section{Shift grids and parameter search}
\label{si:shift_grids}
For Can-CIS(2), both prescriptions used the grid $\varepsilon=0,0.05,\ldots,0.30~E_{\rm h}$, supplemented by $0.11,0.12,\ldots,0.19~E_{\rm h}$. For Hyb-CIS(2), the N-shift used the same coarse grid, supplemented by $0.01,0.02,\ldots,0.10~E_{\rm h}$. For the I-shift, the grid comprised $0,0.01,0.02,\ldots,0.10~E_{\rm h}$ together with $0.025$, $0.075$, $0.15$, $0.20$, and $0.30~E_{\rm h}$. All four prescriptions cover the same states.

Signed errors were interpolated using the piecewise cubic Hermite interpolating polynomial (PCHIP), which preserves monotonicity between consecutive data points, before taking absolute values.\cite{FritschButland1984} The parameter search used a $0.005~E_{\rm h}$ mesh with the selection criteria specified in the main text; the analyses of Sec.~\ref{si:nn_detail} use a $0.001~E_{\rm h}$ mesh for both shifts. Direct calculations at the common parameters, $0.125~E_{\rm h}$ for the Canonical I-shift and $0.035~E_{\rm h}$ for the Hybrid N-shift, were added to the grids. The main-text prescriptions and the I-shift analyses of this Supporting Information use these grids. For the supplementary N-shift analyses, the original grid was extended to $-0.15$ through $-0.05~E_{\rm h}$ and $0.35$ through $0.45~E_{\rm h}$ for states whose signed errors crossed zero outside the initial range, and refined to a spacing of $0.0125~E_{\rm h}$ between $0.125$ and $0.25~E_{\rm h}$.

\section{Dependence on the shift parameter}
\label{sec:linear}

\begin{figure*}[t]
\includegraphics[width=0.92\textwidth]{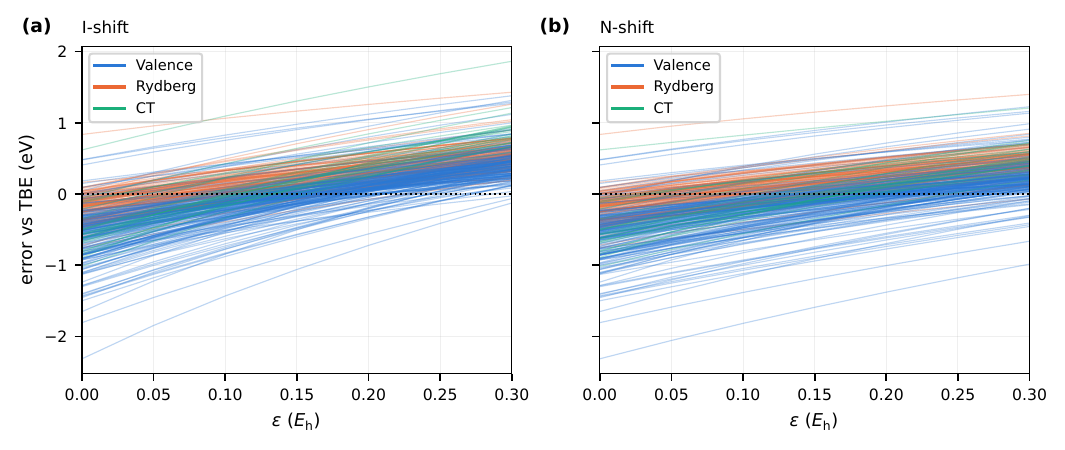}
\caption{Response of Can-CIS(2) to the shifts. The error against the TBE is plotted as a function of $\varepsilon$ for each state with (a) the I-shift and (b) the N-shift.}
\label{fig:si_linN}
\end{figure*}

The excitation energies of Can-CIS(2) depend nearly linearly on the parameter of both shifts [Figure~\ref{fig:si_linN}]. For the I-shift, the median coefficient of determination of a linear fit is $0.996$, and the median slopes are $3.17$, $2.40$, and $4.71$~eV per $E_{\rm h}$ for the valence, Rydberg, and CT states. This linear dependence underlies the estimate of $\varepsilon^{\rm opt}$ in Figure~5(a) of the main text, and the larger slope of the CT states accounts for the steeper dependence of their MAE on $\varepsilon$.

\section{Additional analyses of the I- and N-shifts}
\label{si:nn_detail}
\label{si:cisd_shift}

This section compares Can-CIS(2) and Hyb-CIS(2) with the I- and N-shifts. The common parameters are those of the main text, determined on a $0.005~E_{\rm h}$ mesh. The optimal parameters and the criterion comparison below use a $0.001~E_{\rm h}$ mesh for both shifts, and the N-shift results use the extended grids of Sec.~\ref{si:shift_grids}. V and R denote the valence and Rydberg states of QUEST\#1 and \#3, and CT denotes the 24 charge-transfer states.

\subsection{Accuracy with a common parameter}

Table~\ref{tab:si_cisd_main} gives the MAEs for each benchmark set and class for the two CIS(2) partitions. The common parameters are $\varepsilon^{\mathrm I}=0.125$ and $\varepsilon^{\mathrm N}=0.160~E_{\rm h}$ for Can-CIS(2) and $0.030$ and $0.035~E_{\rm h}$ for Hyb-CIS(2).

\begin{table}[htbp]
\caption{Accuracy of the I- and N-shifts for CIS(2). MAEs (eV) without a shift (Unshifted) and with the I- and N-shifts at the common parameters, $\varepsilon^{\mathrm I}=0.125$ and $\varepsilon^{\mathrm N}=0.160~E_{\rm h}$ for Can-CIS(2) and $0.030$ and $0.035~E_{\rm h}$ for Hyb-CIS(2). V and R denote the valence and Rydberg states of QUEST\#1 and \#3, and CT the 24 charge-transfer states.}
\label{tab:si_cisd_main}
\begin{ruledtabular}
\begin{tabular}{lrcccccc}
 & & \multicolumn{3}{c}{Can-CIS(2)} & \multicolumn{3}{c}{Hyb-CIS(2)} \\
\cline{3-5}\cline{6-8}
Subset & $n$ & Unshifted & I & N & Unshifted & I & N \\
\hline
QUEST\#1 singlets & 53 & 0.340 & 0.166 & 0.179 & 0.244 & 0.240 & 0.240 \\
QUEST\#1 triplets & 47 & 0.290 & 0.112 & 0.118 & 0.150 & 0.144 & 0.145 \\
QUEST\#3 singlets & 123 & 0.536 & 0.246 & 0.278 & 0.284 & 0.249 & 0.255 \\
QUEST\#3 triplets & 88 & 0.516 & 0.196 & 0.221 & 0.193 & 0.153 & 0.156 \\
QUEST\#6 singlets & 23 & 0.561 & 0.201 & 0.178 & 0.289 & 0.254 & 0.253 \\
\hline
All & 334 & 0.467 & 0.198 & 0.218 & 0.235 & 0.208 & 0.211 \\
\hline
V & 206 & 0.588 & 0.198 & 0.221 & 0.271 & 0.204 & 0.209 \\
R & 104 & 0.208 & 0.199 & 0.222 & 0.154 & 0.205 & 0.207 \\
CT & 24 & 0.544 & 0.197 & 0.177 & 0.277 & 0.246 & 0.245 \\
\end{tabular}
\end{ruledtabular}
\end{table}
\subsection{Optimal parameters for each benchmark set}
\label{si:nn_optimal}

Table~\ref{tab:si_cisd_opt} lists the parameters that minimize the MAE of each benchmark set separately, together with the resulting MAEs.

\begin{table}[htbp]

\caption{Optimal shift parameters for each benchmark set. Optimal parameters $\varepsilon^\star$ ($E_{\rm h}$, searched on a mesh of $0.001~E_{\rm h}$) of the I- and N-shifts and the MAEs (eV) at these parameters for each set.}
\label{tab:si_cisd_opt}
\begin{ruledtabular}
\begin{tabular}{lrcccc}
 & & \multicolumn{2}{c}{Can-CIS(2)} & \multicolumn{2}{c}{Hyb-CIS(2)} \\
\cline{3-4}\cline{5-6}
Subset & $n$ & $\varepsilon^\star$ & MAE & $\varepsilon^\star$ & MAE \\
\hline
\multicolumn{6}{l}{I shift}\\
QUEST\#1 singlets & 53 & 0.108 & 0.163 & 0.031 & 0.240 \\
QUEST\#1 triplets & 47 & 0.133 & 0.111 & 0.022 & 0.143 \\
QUEST\#3 singlets & 123 & 0.138 & 0.242 & 0.048 & 0.240 \\
QUEST\#3 triplets & 88 & 0.148 & 0.183 & 0.040 & 0.150 \\
QUEST\#6 singlets & 23 & 0.109 & 0.185 & 0.028 & 0.254 \\
All & 334 & 0.133 & 0.197 & 0.040 & 0.206 \\
\hline
\multicolumn{6}{l}{N shift}\\
QUEST\#1 singlets & 53 & 0.122 & 0.170 & 0.036 & 0.240 \\
QUEST\#1 triplets & 47 & 0.160 & 0.118 & 0.024 & 0.144 \\
QUEST\#3 singlets & 123 & 0.171 & 0.277 & 0.056 & 0.245 \\
QUEST\#3 triplets & 88 & 0.191 & 0.211 & 0.048 & 0.154 \\
QUEST\#6 singlets & 23 & 0.153 & 0.177 & 0.043 & 0.251 \\
All & 334 & 0.166 & 0.218 & 0.049 & 0.208 \\
\end{tabular}
\end{ruledtabular}
\end{table}

To assess sensitivity to the fitting criterion, we compare three choices. For Can-CIS(2) with the I-shift, minimizing the larger of the V and R MAEs of QUEST\#1 and \#3 (the criterion of the main text) gives $\varepsilon^\star=0.125$, weighting the three classes equally gives $0.110$, and weighting all states equally gives $0.133$; the common value $0.125$ lies within $0$, $3$, and $2$~meV of the minimum of each criterion. For the Hybrid partition with the I-shift, the three criteria give $0.030$, $0.028$, and $0.040$, and $0.030$ lies within $2.2$~meV of each. On the extended grids, the same criteria give $0.159$, $0.150$, and $0.166$ for the Canonical N-shift and $0.036$, $0.029$, and $0.049$ for the Hybrid N-shift, with $0.160$ and $0.035$ within $2.6$~meV of each.

Both partitions require a larger parameter for valence than for Rydberg states with both shifts. At the optima for each class (Table~V of the main text), Can-CIS(2) retains the lowest CT MAE, so its CT advantage over Hyb-CIS(2) cannot be attributed solely to the choice of the common parameter.

For Hyb-CIS(2), direct calculations at the common parameters give overall MAEs of $0.2077$~eV for the I-shift and $0.2109$~eV for the N-shift, in agreement with the interpolated values.

Using the overall MAE as the fitting criterion, we further examine the stability of the Can-CIS(2) parameter. Fitting singlets and triplets separately gives $0.129$ and $0.146$ for the I-shift and $0.159$ and $0.178$ for the N-shift; the resulting I/N MAEs are $0.219/0.240$~eV for singlets and $0.159/0.180$~eV for triplets, so the advantage of the I-shift is not confined to one spin multiplicity. Removing one molecule at a time and refitting changes the optimum only between $0.129$ and $0.134$ for the I-shift and between $0.160$ and $0.168$ for the N-shift over the 60 molecules. The optima of individual molecules vary more widely: for the 41 molecules with at least three states, the median and interquartile range are $0.125$ and $0.087$--$0.152$ for the I-shift and $0.151$ and $0.090$--$0.191$ for the N-shift. For both shifts, the lowest values are found for molecules whose low-lying states are Rydberg states and the highest for aza-aromatic molecules with several valence $n\to\pi^*$ states, again reflecting the class dependence.

\section{Remaining errors}
\label{sec:residualstates}

For Can-CIS(2) with the I-shift at $\varepsilon=0.125~E_{\rm h}$, 130 of the 334 states are within $0.1$~eV of the reference and 264 within $0.3$~eV (N-shift at $0.160~E_{\rm h}$: 119 and 245), so that a minority of states contributes much of the MAE. The ten largest errors fall into two groups. Five are underestimated valence states of tetrazine and triazine, including the $n\to\pi^*$ states tetrazine $2\,^1B_{1g}$ ($-1.24$~eV), tetrazine $2\,^3A_u$ ($-0.98$~eV), and triazine $1\,^1A_1''$ ($-0.74$~eV), which have several close-lying states of the same symmetry; the shift reduces but does not remove their errors, and the CIS reference is the least reliable for them. The other five are already overestimated without the shift: the charge-transfer state of the dipeptide ($+1.20$~eV), the Rydberg state of dinitrogen ($+1.11$~eV), the $1\,^3A_2$ $n\to\pi^*$ state of cyclopropenone ($+0.90$~eV), butadiene $2\,^1A_g$ ($+0.86$~eV), whose multiconfigurational character cannot be described by a CIS reference, and the $1\,^1B_2$ $\pi\to\pi^*$ state of pyrrole ($+0.83$~eV). For these states, the positive shift raises the excitation energy and increases the error. These states expose limitations of a single nonnegative shift and of the CIS reference.

\section{Data files}
\label{si:states}
The data are provided in the spreadsheet \path{SI_data.xlsx}, one table per sheet; its first sheet describes the tables.

\end{document}